\documentclass[aps,prf,preprint,groupedaddress,longbibliography]{revtex4-1}
\usepackage{graphicx} 
\usepackage{color}
\usepackage{times,mathptmx}
\usepackage{graphicx} 	
\usepackage{amsmath,amssymb}
\usepackage[normalem]{ulem}
\usepackage{cancel}

\newcommand{\dd}{\mathrm{d}}

\newcommand{\ra}{\mathrm{Ra}}
\newcommand{\sch}{\mathrm{Sc}}

\begin{document}

\title{Role of solutal free convection  on interdiffusion \\
in a horizontal microfluidic channel}
  
\author{Jean-Baptiste Salmon}
%\email[]{jean-baptiste.salmon-exterieur@solvay.com}
\affiliation{CNRS, Solvay, LOF, UMR 5258, Univ. Bordeaux, F-33600 Pessac, France.}
\author{Laurent Soucasse}
%\email[]{jean-baptiste.salmon-exterieur@solvay.com}
\affiliation{Laboratoire EM2C, CNRS, CentraleSup\'elec, Universit\'e Paris-Saclay, Grande Voie des Vignes, 92295 Chatenay-Malabry cedex, France.}

\author{Fr\'ed\'eric Doumenc}
\affiliation{Universit\'e Paris-Saclay, CNRS, FAST, 91405, Orsay, France,}
\affiliation{Sorbonne Universit\'e, UFR 919, 4 place Jussieu, F-75252, Paris Cedex 05, France.}

\date{\today}

\begin{abstract}
We theoretically investigate the role of solutal free convection on the diffusion of a buoyant solute at the microfluidic scales, $\simeq 5$--$500~\mu$m. We first consider a horizontal microfluidic slit, one half of which initially filled with a binary solution (solute and solvent), and the other half with pure solvent. 
The buoyant forces generate a gravity current that couples to the diffusion of the solute. We perform numerical resolutions of the 2D model describing the transport of the solute in the slit. This study allows us to highlight different regimes  as a function of a single parameter, the Rayleigh number $\text{Ra}$ which compares gravity-induced advection  to solute diffusion. We then derive asymptotic analytical solutions to quantify the width of the mixing zone as a function of time in each regime and establish a diagram  that makes it possible to identify the range of $\text{Ra}$  and times  for which buoyancy does not impact diffusion.
In a second step, we present numerical resolutions of the same  model but for a 3D microfluidic channel with a square cross-section. 
We observe the same regimes as in the 2D case, and focus on the dispersion regime at long time scales. We then derive the expression of the 1D dispersion coefficient for a channel with a rectangular section, and analyse the role of the transverse flow in the particular case of a square section. Finally, we show that the impact of this transverse flow on the solute transport can be neglected for most of the microfluidic experimental configurations.    
\end{abstract}

%\keywords{}

\maketitle

\section{Introduction}
Microfluidics refers to a wide range of technological tools for manipulating liquids in microfabricated networks of channels with cross-sectional dimensions ranging from a few microns to a few hundred microns. Applications of this technology are numerous and diverse, from high throughput miniaturized bioassays to fundamental studies in physical-chemistry, see Refs.~\cite{Convery2019,Whitesides:06,Beebe2002} for some reviews.    
%%%%%%%%%%%%%%%%%%%%%%%%% 	
\begin{figure}[ht]
\begin{centering}
\includegraphics{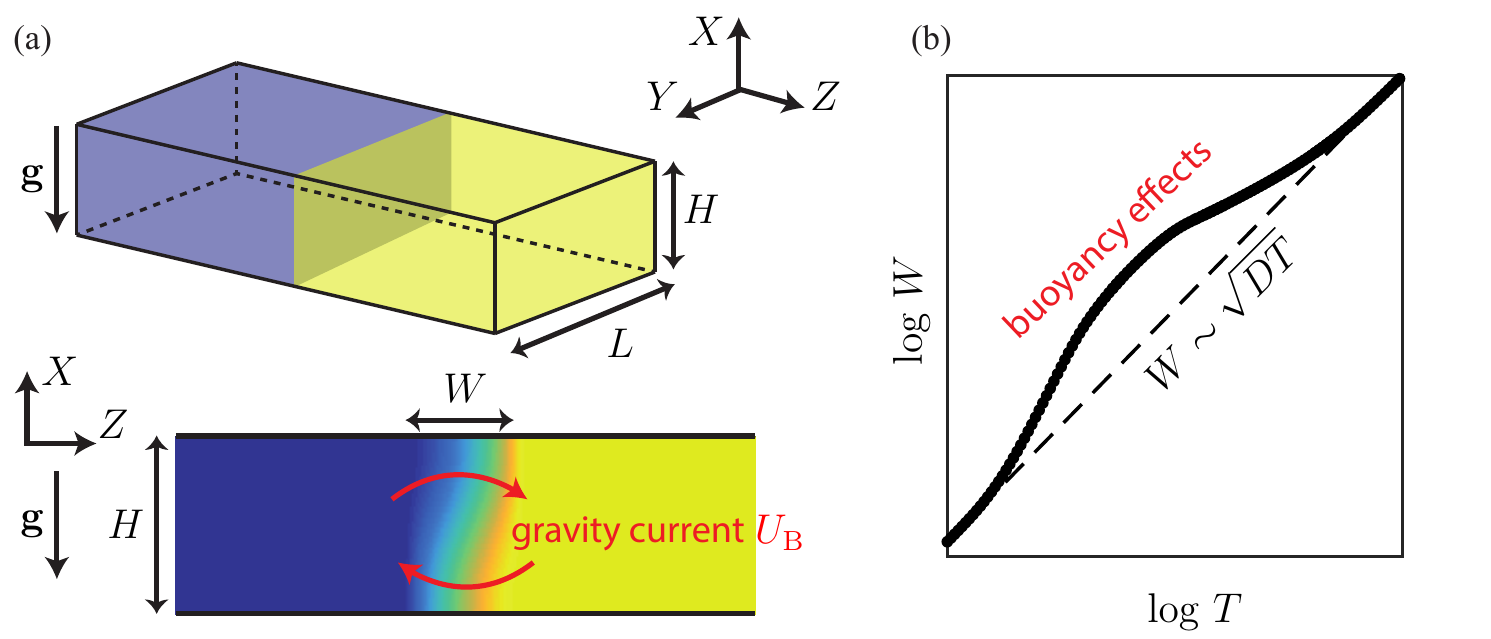}
\caption{(a) Schematic diagram of the diffusion experiment in a microfluidic channel of height~$H$ and width~$L$. The red arrows represent the gravity current $U_\mathrm{B}$
 induced by the difference in density. The color code indicates the concentration of the solute.
(b)  Width of the mixing zone  $W$ vs.\ time $T$. The dashed line is the case for which buoyancy does not affect the spreading of the solute. \label{fig1}}
\end{centering}
\end{figure}
%%%%%%%%%%%%%%%%%%%%%%% 
The small scales of microfluidic technologies allow to study numerous processes while controlling finely  all transport phenomena (mass, momentum, energy)~\cite{Stone2004}.
 In particular, the role of buoyancy has been mentioned by Squires and Quake's in their review on the physics of fluids at the nanoliter scale and quantified using scaling arguments as explained below~\cite{Squires:05}.  
In this reference, the authors considered the situation illustrated in Fig.~\ref{fig1}(a): a  microfluidic channel of height $H$, one half of which is initially filled with a binary solution, solvent and solute at concentration $\Phi_i$, the other half only by the solvent. 
With no difference in density, solute and solvent interdiffuse, and the width of the mixing zone evolves as $W \sim \sqrt{D T}$ where $D$ is the diffusion coefficient of the mixture and 
$T$ the time. 
Now assuming that the density evolves linearly with the  concentration in solute, i.e.:
\begin{eqnarray}
\rho = \rho_0(1 + \beta \Phi)\,, \label{eq:evolrho}
\end{eqnarray}
where $\rho_0$ is the density of the solvent, 
buoyancy induces a gravitational current, the solution flowing under the solvent for $\beta>0$, see Fig.~\ref{fig1}(a).
The magnitude of the gravity current $U_\mathrm{B}$ can be estimated from a balance between buoyant forces $ \sim \rho_0 \beta \Phi_i  g $ and viscous forces $\sim \rho_0 \nu U_\mathrm{B}/H^2$, leading to the velocity scale:
\begin{eqnarray}
U_\mathrm{B} \sim \frac{\beta \Phi_i g H^2}{\nu}\,, 
\end{eqnarray}
where $\nu$ is the kinematic viscosity of the mixture~\cite{Squires:05}.
Note that such a flow exists whatever the height $H$ of the channel as the density gradient is orthogonal to the gravity field $\mathbf{g}$.
The impact of this flow on the solute transport can be determined using the Rayleigh number:
\begin{eqnarray}
  \text{Ra} = \frac{\beta \Phi_i g H^3}{\nu D} \sim \frac{U_\mathrm{B} H}{D}\,,  \label{eq:Ra}
\end{eqnarray}
comparing advection to diffusion. $\ra$ is similar to the P\'eclet number describing mass transport in forced convection problems, but with the important difference that the solute in the present case is not {\it passive},  since the solute itself is the source of the flow~\cite{Squires:05}.
Furthermore, 
a balance between viscous forces $\sim \rho_0 \nu U_\mathrm{B}/H^2$ and inertial forces $\sim \rho_0 U_\mathrm{B}^2/H$  leads to the definition of the Grashof number, 
similarly to the Reynolds number for forced convection:
 \begin{eqnarray}
\text{Gr} = \frac{\beta \Phi_i g H^3}{\nu^2} =  \frac{{\rm Ra}}{{\rm Sc}} \,,  \label{eq:Gr}
\end{eqnarray}
where $\sch = \nu/D$ is the Schmidt number. 
As ${\rm Sc}\geq 10^2$ for most liquid mixtures,  one has thus $\mathrm{Gr} \ll {\rm Ra}$ and viscous dissipation a priori dominates inertia within  	the gravity current in most microfluidic applications~\cite{Squires:05}. 

Many groups have reported gravity-driven currents in microfluidic experiments for which density gradients are imposed either using membranes~\cite{Gu2018} or 
transverse mixing between coflowing miscible liquids~\cite{Yoon:05,Selva:12}. 
In a different context, many groups also reported such flows when density gradients are induced by the evaporation of a liquid mixture in a confined geometry ($H = 100$--$1000~\mu$m) such as sessile drops~\cite{Edwards2018,Li2019,Kang2013}, confined drops~\cite{Pradhan2018,Lee:14,Daubersies:12,Loussert:16}, or micro-capillaries~\cite{Inoue2020,Pradhan2016}. 
The impact of buoyancy on the solute transport is not always mentioned in such works, 
and most groups consider that solutal free convection plays little role at the microfluidic scales although experimental configurations with density gradients are ubiquitous in applications.
 
To illustrate this  point, let us consider interdiffusion between an aqueous NaCl solution at a concentration of 1~M and pure water. For such a mixture, $D \simeq 1.6 \times 10^{-9}$~m$^2$/s, the difference in density is $\simeq 38$~kg/m$^3$ leading to $\beta \Phi_i \simeq  3.8 \times 10^{-2}$, and the kinematic viscosity is $\nu \simeq 10^{-6}~$m$^2$/s~\cite{Pradhan2016}.
For a microfluidic channel of height $H=20~\mu$m, one finds $\text{Ra}\simeq 1.9$, but due to the scaling $\ra \propto H^3$, the Rayleigh number increases to $\text{Ra}\simeq 230$ for $H=100~\mu$m, and even up to $\text{Ra}\simeq 2.9 \times 10^4$ for $H=500~\mu$m. This numerical application illustrates the importance of going beyond the scaling laws presented above to quantitatively predict the range of Rayleigh numbers for which free convection plays only a minor role in a microfluidic configuration.
Furthermore, although these buoyancy-induced flows could have little impact on the solute concentration gradients that generate them, they still do exist, and are able to effectively disperse less mobile species in the fluid mixture, such as macromolecules or  colloids~\cite{Selva:12}. These buoyancy-driven flows 
may also have an influence in protein crystallization experiments~\cite{Savino:96,Pradhan2012,Apostolopoulou2020,Pradhan2020}, for evaluating colloidal diffusio-phoresis induced by solute gradients~\cite{Gu2018}, or even in the context of
biological systems for the motility of microorganisms~\cite{Dunstan2018}.
It is therefore necessary to quantify these flows as a function of the density gradients that generate them for predicting their possible role on other species in the case of complex fluid mixtures.

In the present work, we study in depth the configuration presented in Fig.~\ref{fig1} in a microfluidic context, i.e. height $H$
in the $\simeq 5$--$500~\mu$m range. 
This experimental configuration  has been implemented  many times in microfluidic devices, either using valves or sliding walls for instance, for various applications such as protein crystallization or biochemical assays~\cite{Hansen:02,Yamada2016}.  
Our main goal is to quantitatively delineate the range of Rayleigh numbers for which mixing is impacted by buoyancy in such a configuration, and to 
predict the  laws $W$ vs.\ $T$. 
Configurations similar to that shown in Fig.~\ref{fig1} have been studied for chemical or civil engineering applications and environmental issues that involve length scales $H$ ranging typically  from   $0.1$ to $100$~m. In such cases, commonly referred to in the literature  as the "lock-exchange" problem,
molecular diffusion is negligible and inertial effects are often significant~\cite{Hallez2008,Shin2004,Martin2011,Matson2012,Seon2007}.
Special mention should be made however of the work of Szulczewski and Juanes~\cite{Szulczewski2013}, who studied a situation very similar  to that shown in Fig.~\ref{fig1}(a)
including also molecular diffusion, 
but for a  2D porous layer in the context of geological CO$_2$ sequestration. 
Surprisingly, we are not aware of any work that has studied the microfluidic case where diffusion cannot be neglected, which also motivated this work.
 Since our work is related to microfluidic applications, we have explored  Rayleigh numbers up to $\ra = 10^5$.
These high values are at the limit of most microfluidic dimensions, but can easily be obtained as soon as $H$ reaches the millimeter scale, even for dilute solutions. For example, in the numerical application given previously, water and salty water at 1~M, $\ra = 10^5$ for $H \simeq 750~\mu$m. 
As we subsequently consider liquid mixtures only, the 
smallest Schmidt number we explored is $\sch = 10^2$. Such small values can be observed in the case of the diffusion of small molecules in a low-viscosity solvent, e.g.\ water in acetone~\cite{Tyn1975}.

The present paper is organized as follows. In Sec.~\ref{sec:model}, we present the set of equations modeling the transport of the solute in the configuration shown in Fig.~\ref{fig1}, as well as details about the numerical resolutions. In Sec.~\ref{sec:slit}, we study the case of a 2D slit for the sake of simplicity, i.e.\  
two infinite and parallel plates separated by a thickness $H$.
The numerical data show a rich temporal succession of different regimes of  solute spreading, that can be captured using analytical asymptotic solutions,  and the analogy with the case of a 2D porous layer~\cite{Szulczewski2013} is discussed.  We finally address in Sec.~\ref{sec:3D} the case of a 3D microfluidic channel for which transverse flows  also exist. We finally conclude our work in Sec.~V and insist on its possible implications.

\section{Model and numerical resolution\label{sec:model}}

\subsection{Model and dimensionless variables \label{sec:IIa}}
We consider the situation described in Fig.~\ref{fig1}(a): a straight microfluidic channel of infinite length and rectangular cross-section initially filled with a solution 
at concentration $\Phi_i$ for $Z>0$ and by  pure solvent for $Z<0$.
For the sake of simplicity, we consider that the kinematic viscosity $\nu$ and the interdiffusion coefficient $D$ are constant, and that the density of the solution evolves linearly with the volume fraction in solute, Eq.~(\ref{eq:evolrho}). Because of this linearity, the problem described here can trivially also apply to the interdiffusion between two solutions of different concentrations.  

Assuming an isothermia of the system and the Boussinesq approximation, the equations governing the solute transport and the velocity field $\mathbf{U}$  are:
\begin{eqnarray}
&&\rho_0\left( \frac{\partial \mathbf{U}}{\partial T} + \mathbf{U}.\nabla \mathbf{U}    \right)  = \rho_0\nu\Delta \mathbf{U} - \nabla P + (\rho(\Phi)-\rho_0) \mathbf{g}, \label{eq:stokdim}\\
&&\nabla . \mathbf{U} = 0, \label{eq:contdim}\\
&&\frac{\partial \Phi}{\partial T} + \mathbf{U}.\nabla \Phi = D \Delta \Phi, \label{eq:convdim}
\end{eqnarray}
where $P$ is the pressure deviation from the hydrostatic pressure field for $\Phi=0$.
Boundary conditions at the solid walls are  the no-slip   and the impermeability conditions, $\mathbf{U}=0$  for the velocity field and  $\mathbf{n}.\nabla \Phi=0$ for the concentration field. 
We also impose   $\mathbf{U}(Z \to \pm \infty) = 0$ and $\partial_Z \Phi(Z \to \pm \infty) = 0$ resulting in no pressure-driven flow along the channel (i.e. the gravity current is the only flow).
Initial conditions are given by $\mathbf{U}=0$, $\Phi=\Phi_i$ for $Z>0$ and $\Phi=0$ for $Z<0$.  

To get more insights into the mechanisms of solute  transport,  we define the following dimensionless variables:
\begin{eqnarray}
&& x = X/H,~~y = Y/H,~~z = Z/H,~~t = DT/H^2,~~\gamma = L/H\,,\\
&&\mathbf{u} = (H/D) \mathbf{U},~~p=H^2/(\rho_0 \nu D)P,~~\varphi = \Phi/\Phi_i\,.
\end{eqnarray}
With such definitions, the model given by Eqs.~(\ref{eq:stokdim}-\ref{eq:convdim}) reads now:
\begin{eqnarray}
&&\frac{1}{\text{Sc}} \left(\frac{\partial  \mathbf{u}}{\partial t} + \mathbf{u}.\nabla  \mathbf{u} \right) = \Delta \mathbf{u}  - \nabla p - \text{Ra} \varphi \mathbf{e_x}\,, \label{eq:scstokadim}\\
&&\nabla . \mathbf{u} = 0\,, \label{eq:sccontadim}\\
&&\frac{\partial \varphi}{\partial t}  + \mathbf{u}.\nabla \varphi = \Delta \varphi\,, \label{eq:scconvadim}
\end{eqnarray}
where $\mathbf{e_x}$ is the unit vector along $x$.
With these variables, the initial conditions are:
\begin{eqnarray}
&&\mathbf{u}(x,y,z,t=0)=\mathbf{0}~~{\rm and}~~\varphi(x,y,z,t=0) = \mathcal{H}(z)\,,\label{eq:CI3}
\end{eqnarray}
where  $\mathcal{H}(z)$ is the Heaviside function. Boundary conditions are given by:
\begin{eqnarray}
&&\mathbf{u}=0~~{\rm and}~~\mathbf{n}.\nabla \varphi=0 \label{eq:BC2}
\end{eqnarray}
on the solid walls,
and:
\begin{eqnarray}
\mathbf{u}(z\to \pm \infty) = 0~~{\rm and}~~\partial_z \varphi(z\to \pm \infty) = 0\,. \label{eq:CLinf}
\end{eqnarray}

To estimate the role of buoyancy on the solute   spreading,
we first define the cross-section  averaged concentration profile by:
\begin{eqnarray}
\varphi_0(z,t) = <\varphi> =  \frac{1}{\gamma}\int_ {-\gamma/2}^{\gamma/2} \int_0^1  \varphi(x,y,z,t)\text{d}x \text{d}y\,, \label{eq:defphi03D}
\end{eqnarray}
and the extent of the mixing zone by:
	\begin{eqnarray}
w(t) = \sqrt{\frac{1}{2}\int_{-\infty}^{\infty} {z}^2 \frac{\partial \varphi_0}{\partial z}  \text{d}z}\,. \label{eq:defw}
\end{eqnarray} 
In the case of a neutrally-buoyant solute, $\ra=0$ and the model described above admits the following simple solution~\cite{Crank}:
\begin{eqnarray}
&&\mathbf{u} = 0\,,\\
&&\varphi(x,y,z,t) = \varphi_0(z,t) = \frac{1}{2} \left[1+\text{Erf}\left(\frac{z}{2\sqrt{t}}\right)\right]\,. \label{eq:solErf}
\end{eqnarray}
In this case, the width of the interdiffusion zone is given by $w = \sqrt{t}$, i.e.\ $W =\sqrt{D T}$ with real units. 
This is the classical square-root spreading of the solute due to molecular diffusion.
For $\ra>0$, any deviation from this simple law is a priori a signature of  buoyancy-induced dispersion, see Fig.~\ref{fig1}(b).

\subsection{Numerical resolution\label{sec:3Dsimuldetails}}

Eqs.~(\ref{eq:scstokadim}-\ref{eq:BC2}) have been solved numerically for two distinct geometries:  the 2D case of a microfluidic slit ($\gamma \to \infty$) and the 3D case of a rectangular micro-channel with a square cross-section ($\gamma=1$). In both cases, the boundary conditions Eq.~(\ref{eq:CLinf}) at $z \to \pm \infty$ have been moved to $z=\pm \lambda$ where $\lambda$ is a finite distance such that $\lambda \gg w$.

The 2D numerical simulations have been performed with the commercial software Comsol Multiphysics based on finite elements (Galerkin method).
Time discretization is based on implicit Backward Differentiation Formulas, with an adaptive time stepping. 
Spatial discretization was achieved by a structured mesh of Lagrangian elements, linear for the pressure and quadratic for the other variables. 
The mesh convergence has been thoroughly tested by successive refinements.
Computations were made on a  workstation with 32 Intel Xeon $2.10$~GHz processors and  $250$~GB of RAM. 

The 3D simulations required the use of  in-house made software~\cite{xin02,xin-PCFD08}, specifically optimized for the simulation of free convection in cavities on parallel architectures and based on a multidomain spectral method.
Chebyshev collocation is used for spatial discretization of the three dimensions of space. 
The pressure-flow coupling is ensured by a projection method that forces the velocity divergence-free condition. 
Time integration is performed through a second order temporal scheme combining a Backward Differentiation (BDF2) scheme for the linear terms with an Adams Bashforth extrapolation for the convective terms. 
Domain decomposition along the $z$-horizontal direction is carried out by the Schur complement method for parallelization purposes. 
Each spatial domain is a cube of size $1$ in dimensionless units.
The spatial resolution in the $z$-direction has been increased at the first moments of the simulation in order to capture the stiff concentration gradient and the mesh convergence has been checked by observing the decay of the Chebyshev spectral coefficients.
Computations were made  in a HPC facility, using from 40 to 360 Intel Xeon 2.30~GHz processors.

For both 2D and 3D geometries, simulations
%, the boundary conditions at $z \to \pm \infty$ have been replaced by an impermeable wall with no-slip condition at  $z=\pm L$, where $L$ is a length much larger than the mixing zone.
have been divided into several time intervals in order to adapt the simulation parameters to the temporal evolution of the flow. 
For each new time interval, the length $2 \lambda$ of the spatial domain was extended to take into account the increase of the mixing zone.

\section{The case of a slit \label{sec:slit}}
In  this section, we study the case of a microfluidic slit, and the different fields in Eqs.~(\ref{eq:scstokadim}--\ref{eq:scconvadim})  now depend only on the two variables $x$ and $z$.
We performed numerical simulations for a fixed Schmidt number $\text{Sc}=10^5$, and three different Rayleigh numbers $\text{Ra}=10^3$, $10^4$,  and $10^5$, over a wide range of time scales, from $t=10^{-7}$ to $t \simeq 10^6$. To test the role of the Schmidt number, we also performed numerical resolutions for the same Rayleigh numbers,  $\sch = 10^2$, $10^3$, $10^4$ and $10^5$, and time scales ranging from $t=10^{-7}$ to $t \simeq 5 \times 10^{-3}$ as $\sch$ only plays a role at early time scales, see below.

\subsection{The case $\text{Ra}=10^5$ and $\text{Sc}=10^5$}
\begin{figure}[ht]
\begin{centering}
\includegraphics{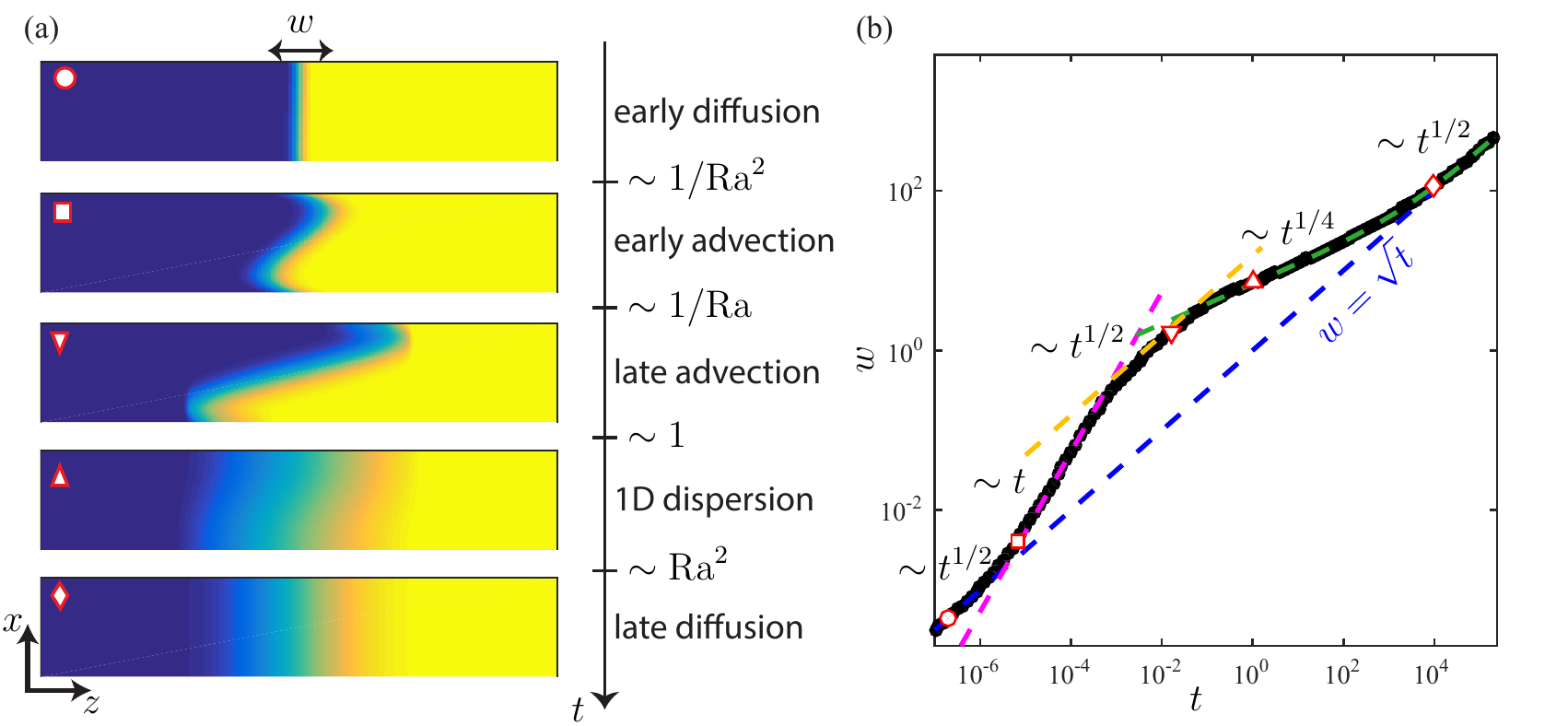}
\caption{For $\text{Ra}=10^5$ and $\text{Sc}=10^5$: (a) concentration fields $\varphi(x,z,t)$ at the times indicated by the corresponding symbols in (b), the $z$-scale is different in each image.
(b) Width of the mixing zone $w(t)$. 
The magenta dashed line is given by Eq.~(\ref{eq:EADw}) and corresponds to early advection with $w\sim t$. The yellow dashed line
is given by Eq.~(\ref{eq:dispLockExchange}) and corresponds to late advection with $w\sim \sqrt{t}$. The green dashed line is computed from the numerical resolution of Eq.~(\ref{eq:diffeff})
and includes both the 1D dispersion regime $w \sim t^{1/4}$ and late diffusion $w\sim \sqrt{t}$.
The blue dashed line is the diffusion law $w = \sqrt{t}$.
See also movie M1 corresponding to these data in the ESI. 
  \label{fig2}}
\end{centering}
\end{figure}

We begin with the case $\text{Ra}=10^5$ and $\text{Sc}=10^5$. Figure~\ref{fig2}(a) displays several 2D concentration maps $\varphi(x,z,t)$ at the times 
shown in Fig.~\ref{fig2}(b),  see also movie M1 corresponding to
these data in the ESI.
%%%%%%%%%%%%%%%%%%%%%%%%% 	
%%%%%%%%%%%%%%%%%%%%%%% 
These data evidence a combination of solute spreading by diffusion and buoyancy-induced advection. 

Figure~\ref{fig2}(b) displays $w(t)$ computed from the numerical resolution for the case  $\ra=10^5$ and $\sch=10^5$, along with the diffusion law $w = \sqrt{t}$. These data clearly show that the mixing zone evolves according to the diffusion law expected without buoyancy at small time scales $t \ll 10^{-5}$, but also at long time scales $t \gg 10^{4}$. For intermediate times, the width of the mixing zone is significantly larger, evidencing the role of the buoyancy-driven advection. The $w$ vs.\ $t$ behavior can be rationalized using different regimes, each with a given power law $w \sim t^\delta$ shown in Fig.~\ref{fig2}(b).
These regimes are presented below in detail,  along with self-similar asymptotic solutions for the concentration
profiles  $\varphi_0(z,t)$ and the corresponding spreading laws $w$ vs.\ $t$.

\subsection{Early diffusion regime \label{sec:earlydiff2D}}
We first analyse the transport of solute at short time scales. 
%%%%%%%%%%%%%%%%%%%%%%%%% 	
\begin{figure}[ht]
\begin{centering}	
\includegraphics{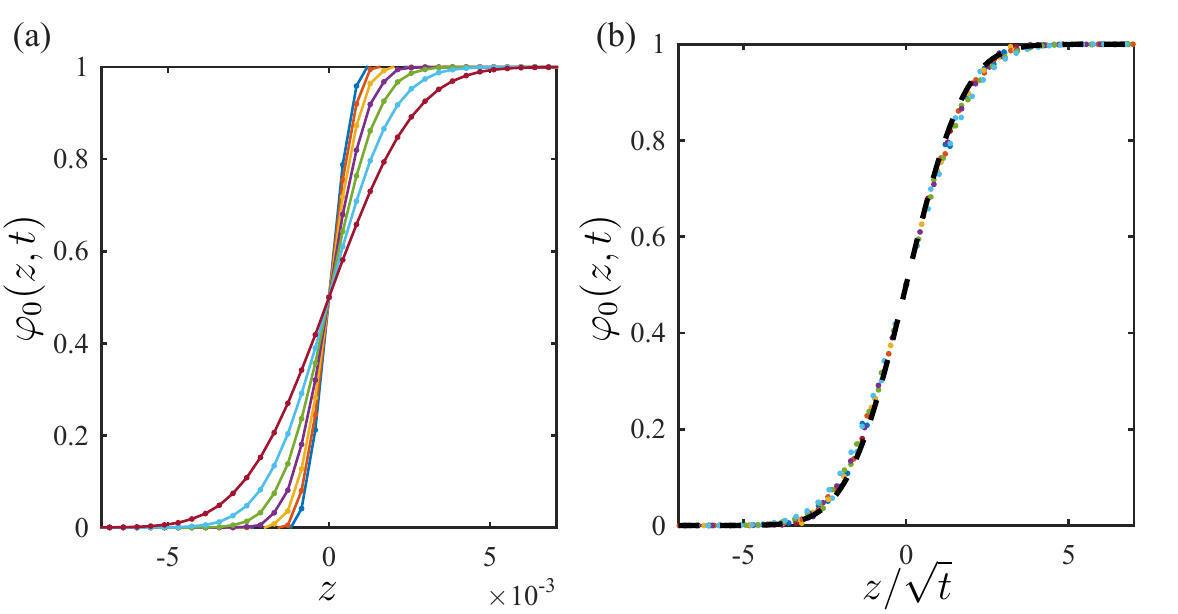}
\caption{(a) Height-averaged concentration profiles $\varphi_0(z,t)$ vs.\ $z$ for time scales ranging from $t = 10^{-7}$ to $t \leq 1.6 \times 10^{-6}$ (7 curves, $\text{Ra}=10^5$ and $\text{Sc}=10^5$). (b) Same data plotted against the reduced variable $z/\sqrt{t}$, the dashed line is given by Eq.~(\ref{eq:solErf}).
\label{fig:earlydiff}}
\end{centering}
\end{figure}
%%%%%%%%%%%%%%%%%%%%%%% 
Figure~\ref{fig:earlydiff}(a) displays the height-averaged concentration profile $\varphi_0(z,t)$ [Eq.~(\ref{eq:defphi03D})] for time scales $t \leq 1.6 \times 10^{-6}$ evidencing the spreading of the solute. As shown in Fig.~\ref{fig:earlydiff}(b), all the profiles collapse on a single curve when plotted against the reduced variable $z/\sqrt{t}$. This curve is correctly described by Eq.~(\ref{eq:solErf}) demonstrating that the 
 transport of the solute is dominated by diffusion for these early time scales, i.e.\ negligible effect of the flow on the solute transport.  As a result, 
the width of the mixing zone computed from the 2D data using Eq.~(\ref{eq:defw}) is correctly fitted by $w = \sqrt{t}$ for $t \leq 1.6 \times 10^{-6}$ as shown in Fig.~\ref{fig2}(b).

\begin{figure}[ht]
\begin{centering}	
\includegraphics{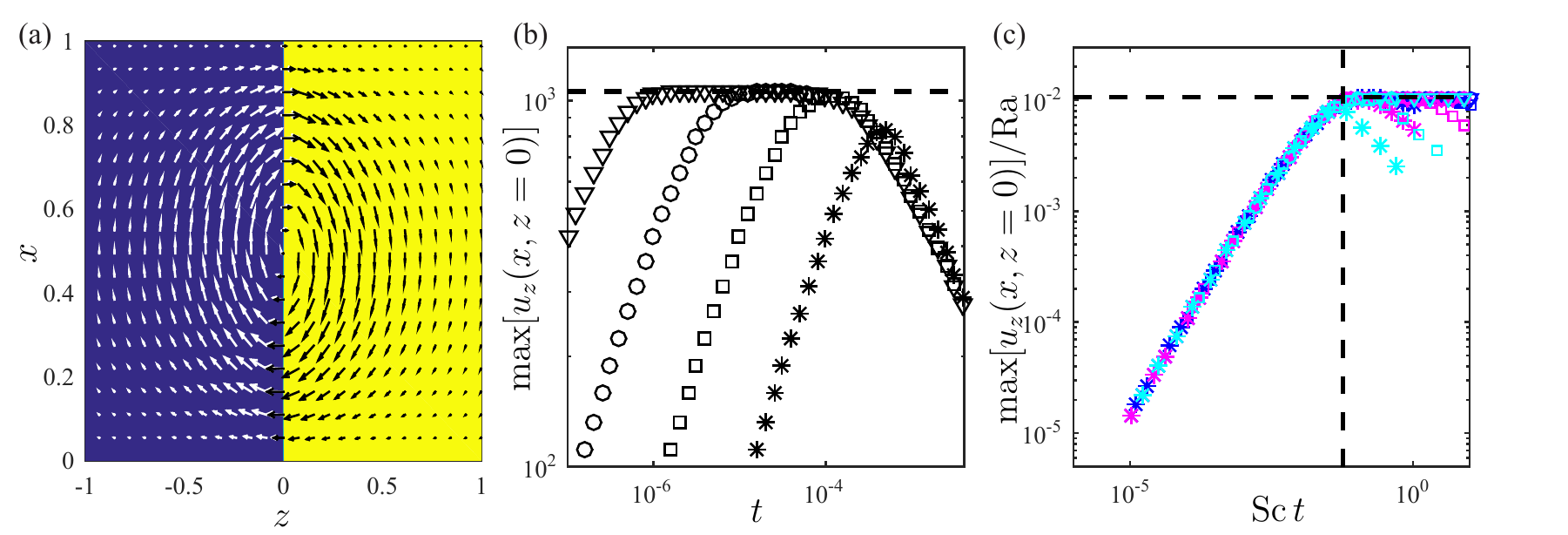}
\caption{(a) Velocity vector field  superimposed with the concentration field at  $t = 1.3 \times 10^{-6}$ ($\ra = 10^5$, $\sch = 10^5$). (b) Maximal value of the component $u_z$  in the plane $z=0$ vs.\ $t$ for $\ra = 10^5$ and $\sch=10^2$ ($\star$), $10^3$ ($\square$), $10^4$ ($\circ$), $10^5$ ($\triangledown$). The horizontal dashed line is $\simeq 1060$. 
(c) Maximal value of  $u_z(x,z=0)$ rescaled by $\ra$ vs.\ $\sch\, t$ for $\ra = 10^5$ (blue), $10^4$ (magenta), and $10^3$ (cyan)
 and $\sch=10^2$ ($\star$), $10^3$ ($\square$), $10^4$ ($\circ$), $10^5$ ($\triangledown$).  The horizontal dashed line is estimated using Eq.~(\ref{eq:uztheoini}) and given by $\simeq 0.0106$, the vertical dashed line is $\sch\,t = 0.0571$.
\label{fig:FigEarlyDiffConv}}
\end{centering}
\end{figure}

However, a flow exists as $\mathbf{u}=0$  does not satisfy the Navier-Stokes equation Eq.~(\ref{eq:scstokadim}), because of the non zero $z$-component of the pressure gradient due to buoyancy. Figure~\ref{fig:FigEarlyDiffConv}(a) indeed displays  the flow driven by this  difference of density for $t \simeq 1.3 \times 10^{-6}$.
This gravity current  corresponds to a recirculating flow developed within the slit on a length scale $\sim 1$.
Figure~\ref{fig:FigEarlyDiffConv}(b) showing the maximal value of the component $u_z$  in the plane $z=0$ in this early  regime, evidences that 
the velocity steadily increases up to reaching a plateau value of $\simeq 10^3$ for time scales $t \geq 0.7 \times 10^{-6}$ for $\sch = 10^5$.
This plot also shows the same data corresponding to several Schmidt numbers, $\sch=10^2$, $10^3$, $10^4$ and $10^5$. After a transient, the 
velocity reaches the same plateau value for all Schmidt numbers, except for $\sch = 10^2$ for which the plateau is not reached.

%%%%%%%%%%%%%%%%%%%%%%%%% 	
%%%%%%%%%%%%%%%%%%%%%%% 

As the transients depend on $\sch$  and thus on the Grashof number $\mathrm{Gr}$, see Eq.~(\ref{eq:Gr}), these results  suggest the existence of an inertial regime corresponding to  the development of the recirculating gravity current through the slit. To better highlight this regime, 
Fig.~\ref{fig:FigEarlyDiffConv}(c) displays the maximal $z$-component of the velocity field at $z=0$ vs.\ time $t$
for  several Rayleigh numbers $\ra= 10^3$, $10^4$, $10^5$ and several Schmidt numbers $\sch=10^2$, $10^3$, $10^4$ and $10^5$.
All the transients collapse on a single curve when times are scaled by $1/\text{Sc}$ and velocities by $\ra$.
This result is recovered from the Navier-Stokes equation Eq.~(\ref{eq:scstokadim}) assuming that the non-linear inertial term $\mathbf{u}.\nabla \mathbf{u}$ does not play any role. The start-up of the flow therefore corresponds simply to the diffusion of the momentum  through the slit, expected to take place on a time scaling as $\sim 1/\text{Sc}$.  By estimating the time it takes for the maximal velocity $u_z$ to reach 90\% of its plateau value, the duration of the inertial regime is $t \simeq 0.0571 / \text{Sc}$,
see the vertical dashed line in Fig.~\ref{fig:FigEarlyDiffConv}(c).
The smallest Schmidt number investigated, $\sch  = 10^2$, is an exception. In this case,
the steady plateau is not observed because inertia is still significant after the end of the diffusion regime, i.e.\ when advection starts affecting the solute transport.
Turning to real units, the duration of the inertial regime  is given by $ \simeq 0.0571\, H^2/\nu$, and lasts only a few tens of milliseconds 
even for $H=500~\mu$m and low-viscosity solvents $\nu = 5\times 10^{-7}$~m$^2$/s. 
This numerical application shows  that such a regime cannot be observed in most microfluidic experiments, as expected,
and that only the Rayleigh number, related to the competition between  gravity-induced advection and diffusion, governs the transport of the solute.

\begin{figure}[ht]
\begin{centering}	
\includegraphics{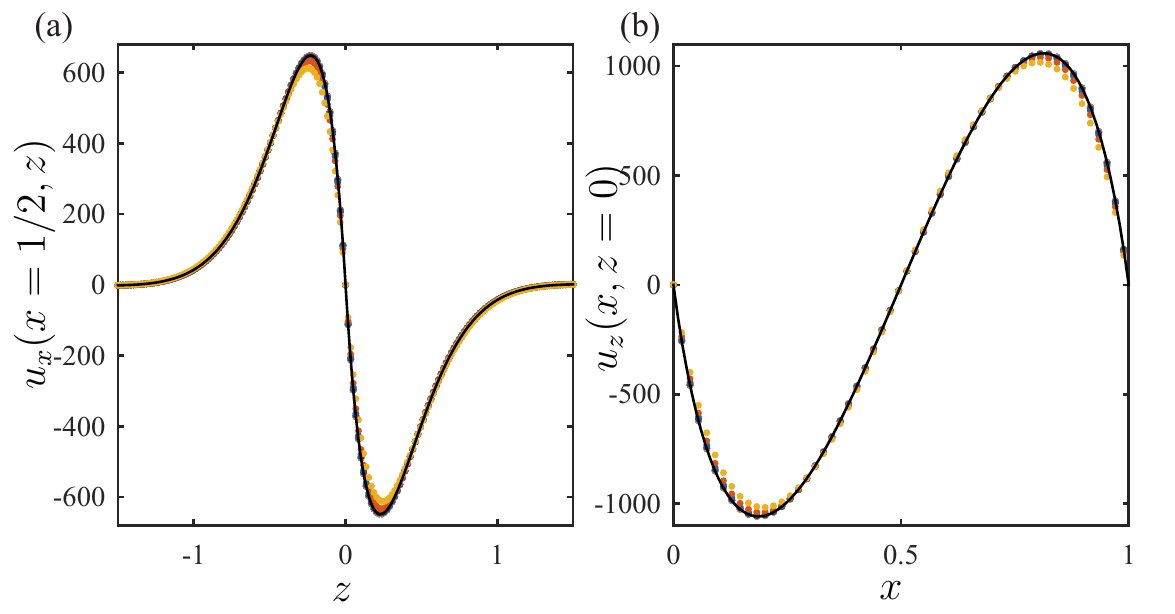}
\caption{(a) $u_x(x=1/2,z)$ vs.\ $z$ and (b)   $u_z(x,z=0)$ vs.\ $x$ for time scales ranging from $t = 1.6 \times 10^{-6}$ to $10^{-4}$ (10 curves, $\ra = 10^5$ and $\sch = 10^5$). The thin black lines are given by Eq.~(\ref{eq:uxtheoini}) in (a) and by 
Eq.~(\ref{eq:uztheoini}) in (b).
\label{fig:earlydiffSuite}}
\end{centering}
\end{figure}

We now turn to the part of the diffusive regime characterized by a steady gravity current, i.e.\  after the end of the inertial transient, see the plateau 
in Fig.~\ref{fig:FigEarlyDiffConv}(b).
For the time scales of this regime, neither diffusion nor advection have significantly widened or distorted the concentration field, and it remains close to the initial condition~Eq.~(\ref{eq:CI3}),  as evidenced by Fig.~\ref{fig:FigEarlyDiffConv}(a) for the case $\ra = 10^5$. 
The velocity field after the inertial transient is therefore expected to be the solution of the steady Stokes equation:
 \begin{eqnarray}
&&0= \Delta \mathbf{u}  - \nabla p - \text{Ra} \mathcal{H}(z) \mathbf{e_x}\,, \label{eq:scstokadimStokes}
\end{eqnarray}	
where $\nabla . \mathbf{u} = 0$ and $\mathcal{H}(z)$ is the Heaviside function.
These equations can be solved analytically, see Appendix~\ref{app:flowearlystage}, leading  to the two following expressions for the components $u_x$ in the plane $x=1/2$ and $u_z$ in the plane $z=0$:
 \begin{eqnarray}
&&u^p_x(x=1/2,z) = \frac{2 \ra}{\pi} \int_{0}^{\infty} \frac{\sinh ^2\left(\frac{k}{4}\right) \left[k-2 \sinh \left(\frac{k}{2}\right)\right]\sin (k z)}{k^3 [k+\sinh (k)]} \, \text{d}k\,, \label{eq:uxtheoini}\\
&&u^p_z(x,z=0) = -\frac{\ra}{\pi}\int_{0}^{\infty} \frac{(x-1) \sinh (k x)+x \sinh (k-k x)}{ k^2 [k+\sinh (k)]} \, \text{d}k\,. \label{eq:uztheoini}
\end{eqnarray}	
These two expressions correctly approximate the velocity profiles in the plateau regime, see the black lines in  Figs.~\ref{fig:earlydiffSuite}(a) 
and~\ref{fig:earlydiffSuite}(b) for the case $\ra = 10^5$ and $\sch = 10^5$.
The maximal velocities are about $\text{max}[u^p_z(x,z=0)] \simeq 0.0106\, \text{Ra}$ [at $x \simeq 0.81$, the plateau value 
in Fig.~\ref{fig:FigEarlyDiffConv}(c)], and  $\text{max}[u^p_x(x=1/2,z)] \simeq 0.0065\, \text{Ra}$  (at $z \simeq -0.23$).

%%%%%%%%%%%%%%%%%%%%%%%%% 	
%%%%%%%%%%%%%%%%%%%%%%% 

\subsection{Early advection \label{sec:earlydiffearlyadv}}

The initial diffusion regime described in Sec.~\ref{sec:earlydiff2D} ceases when the effect of advection on the the solute transport becomes non negligible. 
This transition time corresponds to the departure from the diffusion law $w=\sqrt t$ at $t \simeq 1 \times 10^{-5}$ for the case $\ra=10^5$ and $\sch=10^5$,
see Fig.~\ref{fig2}(b).
A typical concentration field in this new regime is shown in Fig.~\ref{FigAdvection}(a) and has two main characteristics, which can be considered as a definition of the early advection regime:

%%%%%%%%%%%%%%%%%%%%%%%%% 	
\begin{figure}[ht]
\begin{centering} 
\includegraphics{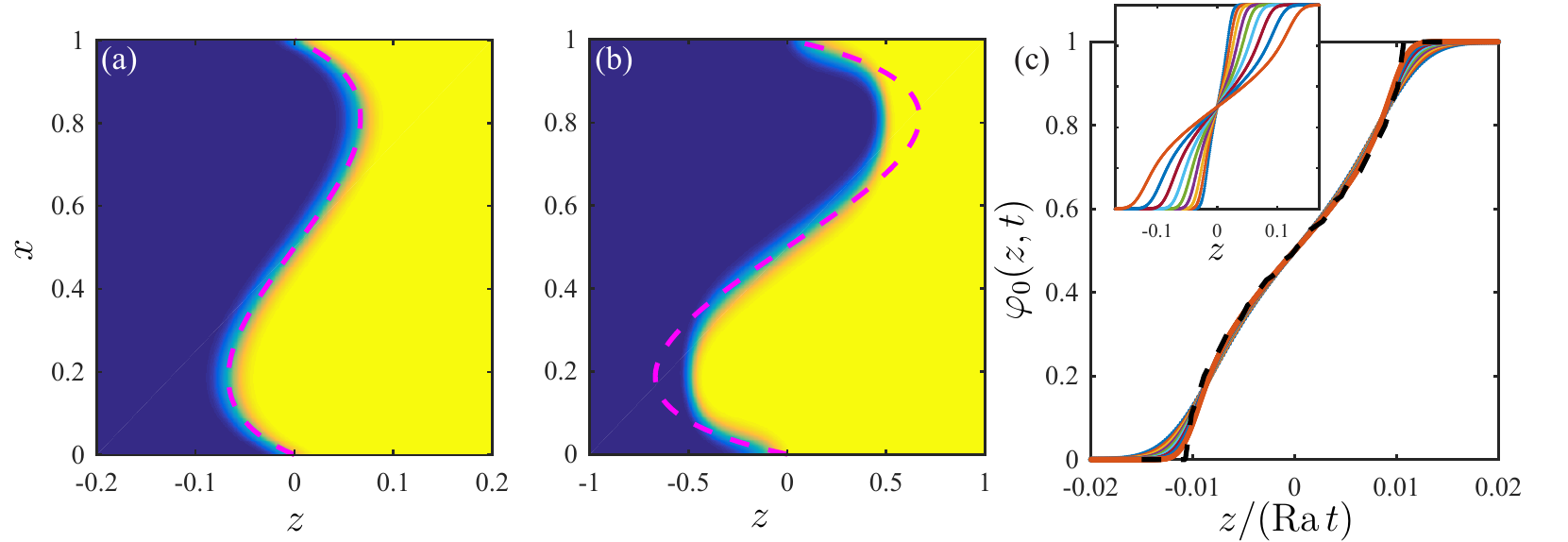}
\caption{Concentration fields $\varphi$ at times (a) $t \simeq 6.3 \times 10^{-5}$ and (b) $t \simeq 6.3 \times 10^{-4}$ for 
$\ra=10^5$ and $\sch=10^5$. The magenta dashed lines are the deformations estimated  by Eq.~(\ref{eq:EADzf}). 
(c) Average concentration profiles against  $z/(\ra\,t)$ for time scales ranging from $t \simeq 2 \times 10^{-5}$ to $1.3 \times 10^{-4}$ (9 curves), the  
dashed line is given by Eq.~(\ref{eq:EADphi2}). Inset: same profiles against $z$. 
 \label{FigAdvection}}
\end{centering}
\end{figure}
%%%%%%%%%%%%%%%%%%%%%%% 

	\paragraph{~} The 2D deformation of the concentration field, obviously due to advection, is much larger than the diffusive spreading. 
For this reason, we can neglect diffusion against advection and approximate the concentration field by an Heaviside function:
\begin{equation} \label{eq:EADphi}
	\varphi(x,z,t)=\mathcal{H}[z-z_f(x,t)] \, ,
\end{equation}
where $z_f(x,t)$ is the position of a front separating two regions, one where $\varphi \simeq 0$ and  the other one where $\varphi \simeq 1$.
		In this advection regime, the front motion along $z$ during a small time interval $\mathrm{d}t$ reads:  
\begin{equation} \label{eq:EADdzf}
	\mathrm{d}z_f(x,t) = u_z(x,z_f(t),t) \, \mathrm{d}t \, .
\end{equation}
		\paragraph{~} The 2D deformation  due to advection is much lower than $1$, i.e.\ than the channel height. 
		The concentration field is thus very close to the initial one, and the velocity field is still given by Eqs.~(\ref{eq:uxtheoini}) and (\ref{eq:uztheoini}) corresponding to the plateau observed in Fig.~\ref{fig:FigEarlyDiffConv}(b). 
		We deduce from these assumptions and Eq.~(\ref{eq:EADdzf}) that the front profile can be approximated by:
\begin{equation} \label{eq:EADzf}
	z_f(x,t) \simeq u_z^p(x,z=0) \, t \,,
\end{equation}
where $u_z^p(x,z=0)$ is the steady velocity field given by Eq.~(\ref{eq:uztheoini}). 
Eq.~(\ref{eq:EADzf}) is consistent with the concentration field obtained from numerical simulations, as shown in Fig.~\ref{FigAdvection}(a).

Eqs.~(\ref{eq:uztheoini}), (\ref{eq:EADphi}) and (\ref{eq:EADzf}) allow the estimation of the height-averaged concentration profile in this regime:
\begin{equation} \label{eq:EADphi2}
%	\varphi_0^{\text{\tiny	{EAR}}}(z,t) = \int_0^1 H[z-  u_z^p(x,z=0)\, t] \, \mathrm{d} x \, .
	\varphi_0(z,t) = \int_0^1 \mathcal{H}[z-  u_z^p(x,z=0)\, t] \, \mathrm{d} x \, .
\end{equation}
One can easily show that the above relation %$\varphi_0^{\text{\tiny	{EAR}}}$ 
is a self-similar function of the variable $z/(\ra \, t)$. 
Figure~\ref{FigAdvection}(c) compares the theoretical relation Eq.~(\ref{eq:EADphi2}) with the numerical simulations, evidencing a reasonably good collapse of the data on the theoretical master curve.
Furthermore, the width of the mixing zone $w(t)$ defined by Eq.~(\ref{eq:defw}) can be estimated using Eq.~(\ref{eq:EADphi2}), leading to:
\begin{equation} \label{eq:EADw}
w \simeq 0.00538 \ra \, t \, .
\end{equation}
This behavior is plotted in Fig.~\ref{fig2}(b) and correctly fits the data obtained from the numerical resolution of the 2D model from $t=1.3 \times 10^{-5}$ to $2 \times 10^{-4}$.

At later time, significant discrepancies are observed between the theoretical formula Eq.~(\ref{eq:EADzf}) and the numerical simulation,  
see for instance Fig.~\ref{FigAdvection}(b) for a comparison at $t=6.3 \times 10^{-4}$. 
Indeed, the deformation of the concentration field is of the order of $1$  for such time scales and Eq.~(\ref{eq:uztheoini}) can no longer be used for the estimation of the velocity field. It marks the end of the early advection regime.

\subsection{Late advection \label{sec:lateadvection}}
\begin{figure}[ht]
\begin{centering}
\includegraphics{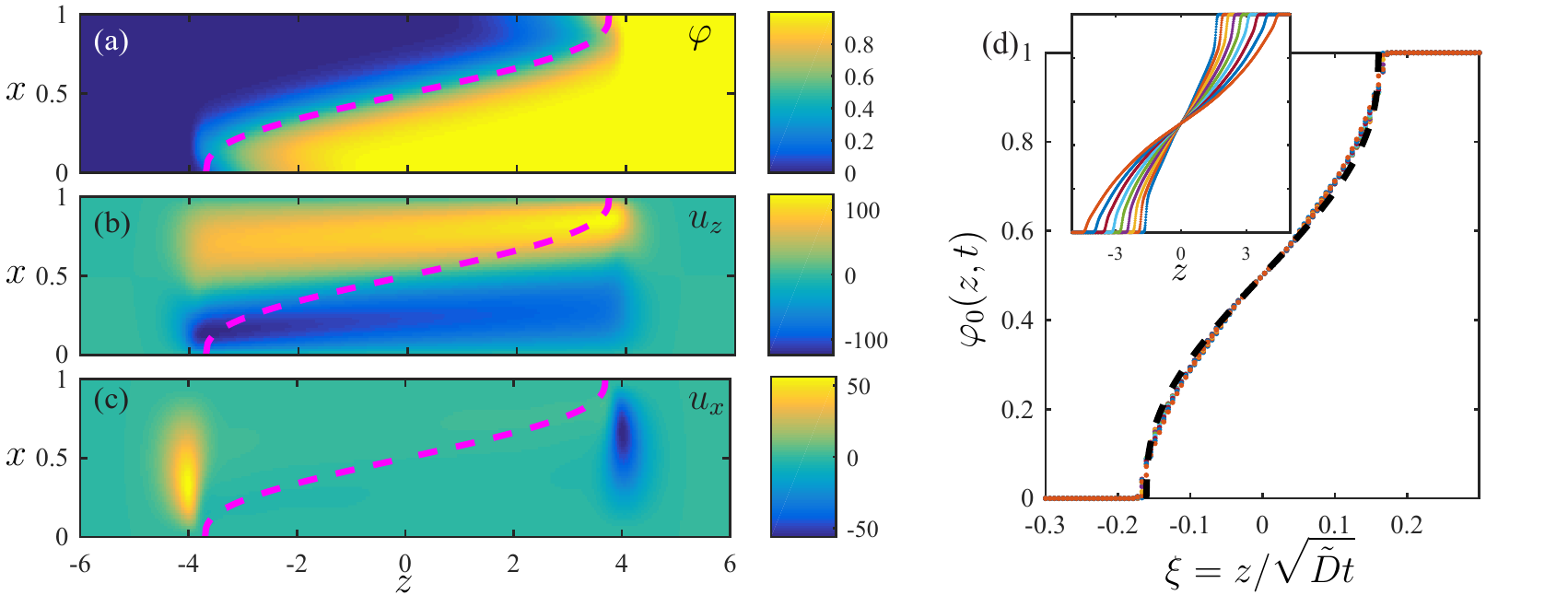}
\caption{(a) Concentration field $\varphi$, and components (b) $u_z$, (c) $u_x$ at  $t  =  1.6 \times 10^{-2}$ for $\ra = 10^5$ and $\sch = 10^5$. 
The magenta dashed line is given by the solution of Eq.~(\ref{eq:Martin}) in the case of a slit. (c) Average concentration profiles $\varphi_0(z,t)$ vs.\ $\xi=z/\sqrt{\tilde{D}t}$ for time scales ranging from  $t \simeq 0.003$ to $0.02$ (9 curves), the dashed line is $\psi(\xi)$ solution of Eq.~(\ref{eq:Martin}). Inset: same profiles against $z$. \label{FigLockExchange}}
\end{centering}
\end{figure}

At later time scales, solute spreading by diffusion still remains negligible as compared to solute advection, but the 2D deformation of the concentration fields is 
now much larger than $1$, see for instance Fig.~\ref{FigLockExchange}(a) showing a snapshot at $t  =  1.6 \times 10^{-2}$. 
%%%%%%%%%%%%%%%%%%%%%%%%% 	
%%%%%%%%%%%%%%%%%%%%%%% 
These data along with the flow field shown in Figs.~\ref{FigLockExchange}(b) and~\ref{FigLockExchange}(c), evidence the reciprocal exchange of the solution and the solvent separated by a diffuse pseudo-interface. For fully negligible  diffusion, this regime commonly referred to as the viscous lock-exchange problem has been widely studied in the literature.
In such a configuration, many groups predicted that the extent of the spreading of the two fluids scales as $w \sim \sqrt{\tilde{D} t}$ where $\tilde{D}$ is an effective diffusion coefficient. The square-root behavior arises from the competition between buoyant forces ($\sim \rho_0 \beta \Phi_i g H/W$) and viscous forces ($\sim \rho_0 \nu \dot{W} / H^2$), leading to the dimensionless scaling law $w^2 \sim \ra \, t$~\cite{Matson2012}.
The  effective diffusion coefficient $\tilde{D}$ therefore only depends on $\ra$ and the geometry, and it has been calculated in various cases: porous medium, circular tube, but also rectangular channel and slit~\cite{Szulczewski2013,Seon2007,Matson2012,Martin2011}.
The main idea of these works is to compute the shape $\psi(z,t)$ of the pseudo-interface separating the two fluids, assuming large deformations and thus a quasi-parallel flow along the $z$-axis (lubrication approximation). With such approximations, one can show  that $\psi(z,t)$  admits a self-similar shape $\psi(\xi)$ with $\xi = z/\sqrt{\tilde{D} t}$, solution of: 
\begin{eqnarray}
-\xi \frac{\dd \psi}{\dd \xi} = 2 \frac{\dd}{\dd \xi}\left( f(\psi) \frac{\dd \psi}{\dd \xi}\right)\,, \label{eq:Martin}
\end{eqnarray}
with $f(\psi) = \psi^3(1-\psi)^3$ and $\tilde{D} = \ra / 3$ for the case of a slit~\cite{Martin2011}.
Equation~(\ref{eq:Martin}) is solved numerically following the method detailed by Martin {\it et al.} in Ref.~\cite{Martin2011}.
Figures~\ref{FigLockExchange}(a--c)  show this solution for the corresponding time $t  =  1.6 \times 10^{-2}$
superimposed with both the concentration field and the velocity field. These data  show  a reasonable agreement, confirming the negligible impact of diffusion on the solute transport.

To better describe this regime, Fig.~\ref{FigLockExchange}(d) displays the average profiles $\varphi_0(z,t)$ obtained from the 2D model at $\ra=10^5$ and $\sch=10^5$ for time scales ranging from $t \simeq 0.003$ to $0.02$. This plot shows that all the data almost collapse on a single curve  when plotted against  $\xi = z/\sqrt{\tilde{D} t}$, which is correctly described by $\psi(\xi)$ solution of Eq.~(\ref{eq:Martin}).
In this regime, one can again compute the width of the mixing zone defined by Eq.~(\ref{eq:defw}) using  $\psi(\xi)$, leading to:
\begin{eqnarray}
w \simeq 0.04879 \sqrt{\ra\,t}\,. \label{eq:dispLockExchange}
\end{eqnarray}
This square-root spreading is plotted in Fig.~\ref{fig2}(b)  and accounts well for the numerical data $w(t)$ obtained for $\ra = 10^5$ and time scales  ranging from $t \simeq 10^{-3}$ to $\simeq 10^{-1}$.
 
\subsection{1D dispersion and late diffusion \label{sec:TaylorDisp}}

%%%%%%%%%%%%%%%%%%%%%%%%% 	
\begin{figure}[ht]
\begin{centering}
\includegraphics{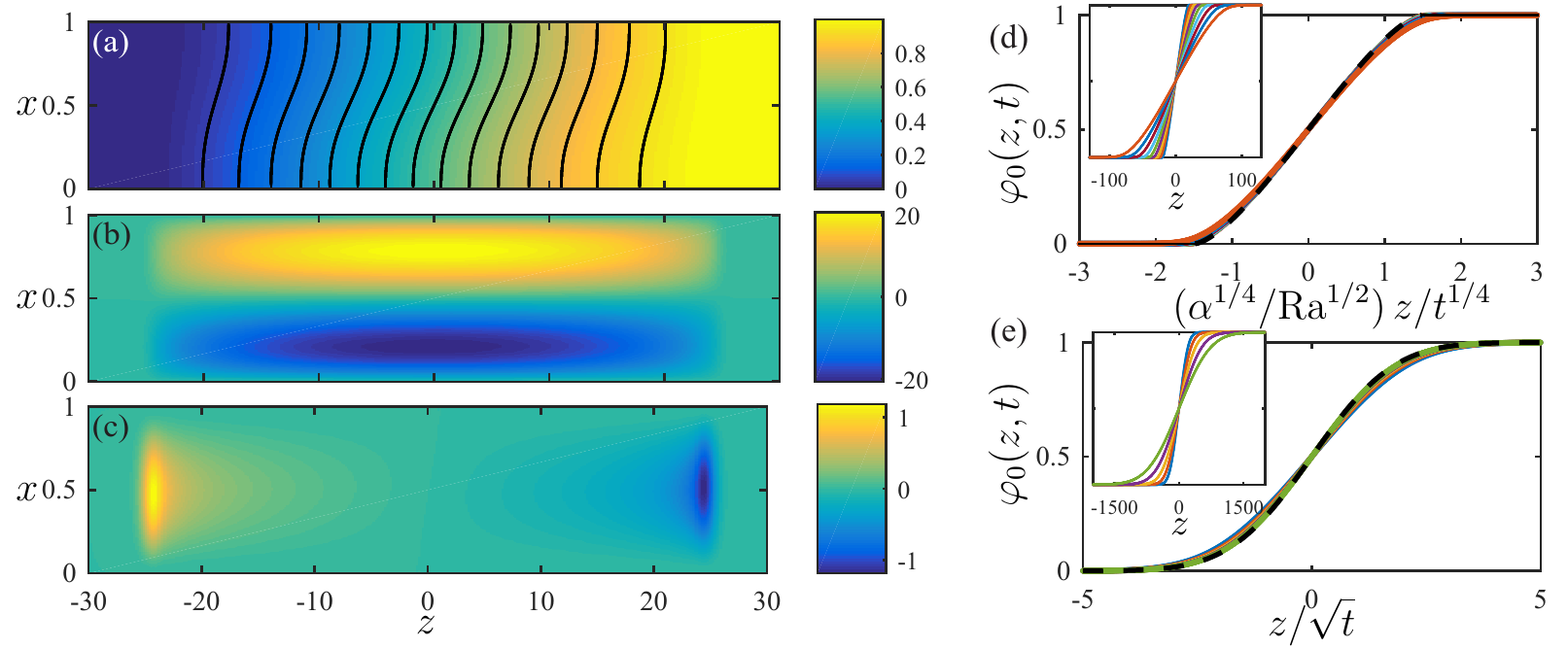}
\caption{(a) Concentration field $\varphi$, (b) component $u_z$ and (c) component $u_x$ at $t=2.5$ ($\ra = 10^5$ and $\sch=10^5$). The black lines in (a) are isoconcentration lines. (d) 
Average concentration profiles $\varphi_0(z,t)$ in the  dispersion regime plotted against $\eta$ given by Eq.~(\ref{diff:eta}), $t$ ranges from $t=1$ to $2.5\times 10^2$ (10 curves). The dashed line is Eq.~(\ref{eq:appAlb}). Inset:
same profiles against  $z$.
 (e) Rescaled concentration profiles $\varphi_0(z,t)$ in the late diffusion regime, $t =  10^4$ up to $1.6 \times 10^5$ (10 curves). The dashed line is Eq.~(\ref{eq:solErf}). Inset:
same profiles against  $z$. \label{FigTaylor}}
\end{centering}
\end{figure}
%%%%%%%%%%%%%%%%%%%%%%% 

For time scales $t \geq \mathcal{O}(1)$, diffusion almost homogenizes the solute over the height of the slit, and  the transport of the solute cannot be described by only advection,
as revealed by Fig.~\ref{FigTaylor}(a) showing  the concentration field at $t=2.5$. In this regime, the  transport   is fully controlled by the coupling between solute  diffusion along the channel height and buoyancy-driven advection along the channel main axis. 
This regime has already been described in the literature since the pioneering work of Chatwin and Erdogan, who studied the Taylor-Aris dispersion of a buoyant solute in a pressure-driven flow~\cite{Erdogan1967}, see also \cite{Smith1976,Barton1976,Godfrey80,MACLEAN:01,Salmon2020} and the review of Young and Jones on shear dispersion~\cite{Young1991}.
In this regime, the extent $w$ of the concentration gradient along $z$ is large ($w \gg 1$), the buoyancy-driven flow is quasi-parallel [$u_x \ll u_z$, see Figs.~\ref{FigTaylor}(b) and~\ref{FigTaylor}(c)], and the variations of the concentration along $x$ are small [see the isoconcentration lines in  Fig.~\ref{FigTaylor}(a)]. One can therefore use the lubrication approximation to show that the density gradient along $z$ drives a flow following:
\begin{eqnarray}
u_z(x,z,t) =  -\frac{\ra}{12} \frac{\partial \varphi_0}{\partial z} x (2 x -1 )(x-1)\,,  \label{eq:compuzTaylor}
\end{eqnarray} 
see Appendix~\ref{app:effdisp2D}. 
This flow  adds a contribution to the dispersion which scales as $\sim u_z^2$,  as for the classical Taylor-Aris dispersion in a Poiseuille flow.
More rigorously, one can demonstrate that the average concentration $\varphi_0(z,t)$ obeys the  1D dispersion equation:
\begin{eqnarray}
&&\frac{\partial \varphi_0}{\partial t} =  \frac{\partial}{\partial z}\left( D_\mathrm{eff} \frac{\partial \varphi_0}{\partial z}\right)\,,
 \label{eq:diffeff}
\end{eqnarray}
with:  
\begin{eqnarray}
D_\mathrm{eff}  = 1 + \frac{1}{\alpha} \left(\text{Ra} \frac{\partial \varphi_0}{\partial z}\right)^2\,, \label{eq:DeffTaylor}
\end{eqnarray}
and $\alpha = 362880$, see Appendix~\ref{app:effdisp2D}. 
The non-linearity of the dispersive term  comes from the coupling between the concentration gradient and the flow, unlike the case of the Taylor-Aris dispersion: strong gradient increases the magnitude of the gravity current which in turn increases the dispersion of the solute.
We will return in detail to the derivation of these equations in the next section when we tackle the 3D case of a microfluidic channel with a rectangular cross-section.

Equation~(\ref{eq:diffeff}) can be made free of any parameter by defining $t^\star = (\alpha/\ra^2)t$ and $z^\star = (\sqrt{\alpha}/\ra)z$. It is
then solved numerically  with the initial condition $\varphi_0(z,t=0) = \mathcal{H}(z)$ to compute the solution for any Rayleigh number and $\alpha$ value. 
Figure~\ref{fig2}(b) shows that the width of the mixing zone $w(t)$ defined by  Eq.~(\ref{eq:defw}) and computed from the numerical resolution of Eq.~(\ref{eq:diffeff}),
perfectly matches the data obtained from the full 2D model for time scales $t\geq 0.1$. 
The component $u_z$ at $z=0$ computed from the 2D model is also very well-approximated by Eq.~(\ref{eq:compuzTaylor}) using the solution of the  1D dispersion model (data not shown). 

Equation~(\ref{eq:diffeff})  along with the initial condition $\varphi_0(z,t=0) = \mathcal{H}(z)$ has been studied by Maclean and Alboussi\`ere~\cite{MACLEAN:01} who 
provided asymptotic  approximations of the solution.
When buoyancy dominates the  transport of the solute, i.e.\ $D_\mathrm{eff} \gg 1$, Eq.~(\ref{eq:diffeff}) admits the 
self-similar solution:
 \begin{eqnarray}
\varphi_0 = \frac{1}{2}+\frac{\eta}{2}\sqrt{\frac{1}{\sqrt{3}\pi} - \frac{\eta^2}{12}} + \frac{1}{\pi}\text{arcsin}\left( \frac{\sqrt{\pi}\eta}{2 \times 3^{1/4}} \right)\,, 
\label{eq:appAlb}
\end{eqnarray}
with $\eta$ given by:
 \begin{eqnarray}
\eta = \frac{\alpha^{1/4}}{\sqrt{{\rm Ra}}} \frac{z} {t^{1/4}}\,. \label{diff:eta}
\end{eqnarray}
Eq.~(\ref{eq:appAlb}) is valid for $\eta^2 \leq 12/(\pi \sqrt{3})$~\cite{MACLEAN:01}.
Figure~\ref{FigTaylor}(d) displays this asymptotic self-similar solution for $\ra = 10^5$ along with the data computed from the 2D model, evidencing a very good agreement.
In this regime, one can compute the width of the mixing zone leading to:
\begin{eqnarray}
w \simeq   \sqrt{\frac{\text{Ra}}{2 \pi}} \left(\frac{3 t}{\alpha}\right)^{1/4}\,, \label{eq:dispTA}
\end{eqnarray}
thus following $w \sim t^{1/4}$ as shown in Fig.~\ref{fig2}(b).

At later times, the density gradient continuously decreases as the solute is continuously dispersed along the channel, and diffusion dominates again the transport of the solute, i.e.\ $D_\mathrm{eff} \simeq 1$. In this late diffusion regime, we once again find a classical diffusion problem where solutal free convection no longer plays a role, and the concentration profiles are then given by Eq.~(\ref{eq:solErf}).
This is illustrated by Fig.~\ref{FigTaylor}(e) showing Eq.~(\ref{eq:solErf}) superimposed with the average concentration profiles computed from the  2D model at long time scales. In this late regime, the width of the mixing zone is again given by $w = \sqrt{t}$ as shown in Fig.~\ref{fig2}(b), despite the buoyancy-driven flow along the slit still given by Eq.~(\ref{eq:compuzTaylor}). As explained in Introduction, even if this flow has no effect on the solute gradient that generates it, it still exists and may have an effect on less mobile species in the case of complex liquid mixtures. 

\subsection{Dispersion diagram}

\begin{table}
\centering
\begin{tabular}{|l|l|l|l|l|}
 \hline
	\hline
 I- early diffusion & $w \simeq \sqrt{t}$ & $\bar{u}_z(z=0,t) \simeq 0.00690\,\ra$ & $t_{\mathrm{I}\to \mathrm{II}} \simeq 34550/\ra^2$\\
	II- early advection & $w \simeq  0.00538\,\text{Ra}\,t$ &$\bar{u}_z(z=0,t) \simeq 0.00690\,\ra$ & $t_{\mathrm{II}\to \mathrm{III}}\simeq 82/\ra $  \\
	III- late advection & $w \simeq 0.04879  \sqrt{\ra\,t}~$ & $\bar{u}_z(z=0,t) \simeq 0.03138\sqrt{\frac{\ra}{t}}$ & $t_{\mathrm{III}\to \mathrm{IV}} \simeq 0.04$ \\
	IV- 1D dispersion& $w \simeq  \sqrt{\frac{\text{Ra}}{2 \pi}} \left(\frac{3 t}{\alpha}\right)^{1/4}$ &$\bar{u}_z(z=0,t)  \simeq \frac{\sqrt{\ra}}{192 \sqrt{\pi}} \left(\frac{\alpha}{3 t}\right)^{1/4} $  & $t_{\mathrm{IV}\to \mathrm{V}} \simeq 
	\frac{3}{(2\pi)^2\alpha}\,\ra^2$ \\
	V- late diffusion& $w \simeq \sqrt{t}$ & $\bar{u}_z(z=0,t) \simeq  \frac{1}{384 \sqrt{\pi}}\,\ \frac{\ra}{\sqrt{t}}$& ~\\
	\hline
  \hline
\end{tabular}
\caption{Transport  regimes and transition times for a slit, $\alpha = 362880$.}
\label{tab:transition}
\end{table}

In the previous paragraphs, we identified a sequence of regimes of  transport of the solute, which are summarized in Table~\ref{tab:transition}.
For each regime, one can compute the typical longitudinal velocity at $z=0$  defined by:
\begin{eqnarray}
\bar{u}_z(z=0,t) = -2 \int_0^{1/2} u_z(x,z=0,t) \text{d}x, \label{eq:defuz}
\end{eqnarray} 
using in particular Eq.~(\ref{eq:uztheoini}) for Regimes~I and II, Ref.~\cite{Martin2011} for Regime~III, and Eq.~(\ref{eq:compuzTaylor}) along with  Eq.~(\ref{eq:appAlb})  
[resp. Eq.~(\ref{eq:solErf})] for Regime~IV (resp. Regime~V). Results are displayed in the third column of  Table~\ref{tab:transition}.

\begin{figure}[ht]
\begin{centering}
\includegraphics{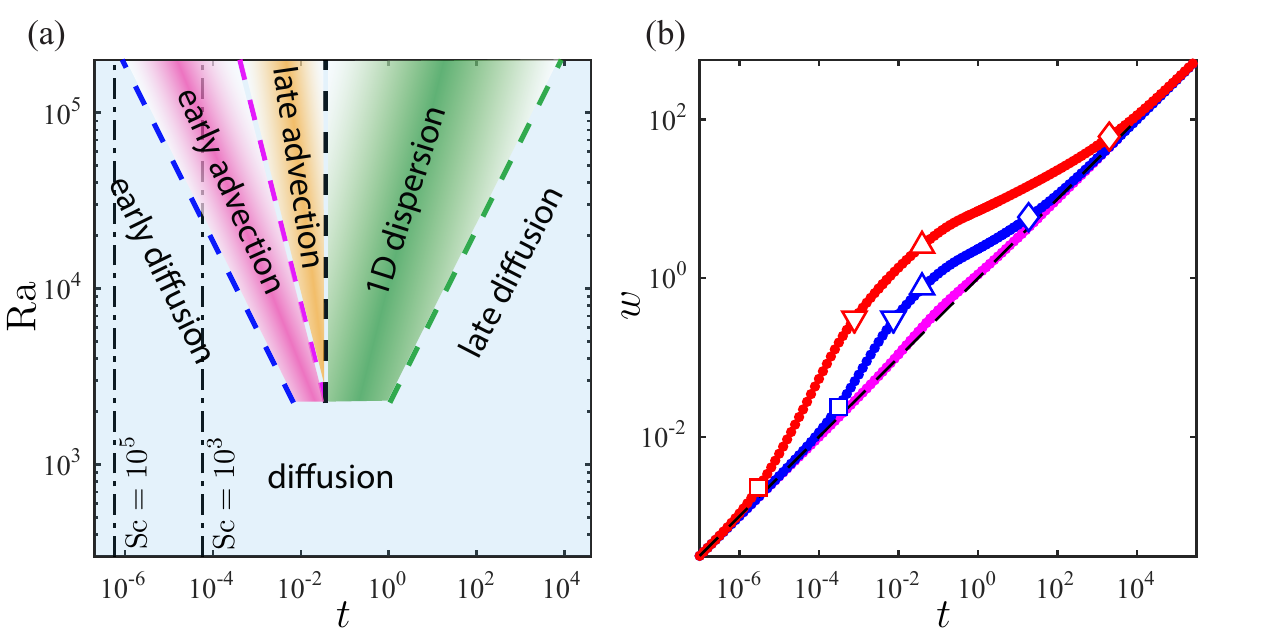}
\caption{(a) Diagram of the different regimes in the plane $\ra$ vs.\ $t$. The transition times $t$ vs.\ $\ra$ are given in Table~\ref{tab:transition}.
The vertical dotted lines correspond to the extent of the inertial regime $t \simeq 0.0571 / \text{Sc}$ for $\sch = 10^3$ and $\sch = 10^5$, see Sec.~\ref{sec:earlydiff2D}.
(b) Width of the mixing zone $w(t)$ for $\sch = 10^5$ and $\ra = 10^5$ (red), $10^4$ (blue), $10^3$ (magenta). The black dashed line is the purely diffusive spreading $w = \sqrt{t}$. The symbols are the transition times given in Table~\ref{tab:transition} for  $\ra = 10^5$ and $\ra = 10^4$.  
  \label{FigDiagramme}}
\end{centering}
\end{figure}

The transition times between these  regimes are estimated by matching the different spreading laws  $w$ vs.\ $t$ leading to the values provided in  Table~\ref{tab:transition}. These definitions lead to 
different scaling laws with $\ra$ which are also reported in Fig.~\ref{fig2}. 
These transition times allow us to construct the diagram presented in Fig.~\ref{FigDiagramme}(a) in the plane $\ra$ vs.\ $t$.
The vertical dashed-dotted lines in Fig.~\ref{FigDiagramme}(a) corresponding to the extent of the inertial regime show that the
early advection regime II is also coupled for large $\ra$ and small $\sch$ to the momentum diffusion across the slit, see 
Sec.~\ref{sec:earlydiff2D} and in particular Fig.~\ref{fig:FigEarlyDiffConv}(b).
%%%%%%%%%%%%%%%%%%%%%%%%% 	
%%%%%%%%%%%%%%%%%%%%%%% 

This diagram also shows that the effect of buoyancy vanishes  for $\ra \lesssim 10^3$.
To illustrate this point, Fig.~\ref{FigDiagramme}(b) reports the  spreading  laws $w$ vs.\ $t$ computed  from the numerical resolution of the full 2D model for several Rayleigh numbers $\ra = 10^5$, $10^4$, $10^3$ and $\sch = 10^5$. These data clearly reveal that buoyancy has little effect on the transport of the solute at all time scales for $\ra = 10^3$.	
More quantitatively, the ratio $w/\sqrt{t}$ reaches a maximum of only $\simeq 1.24$ at $t \simeq 0.1$ for $\ra = 10^3$. 
This result answers the question initially asked in Introduction as it allows to assess a numerical value to $\ra$ corresponding to negligible buoyancy in a microfluidic slit at all time scales,  at least regarding the {\it active} solute that generates the gravity current.

The regimes of early diffusion and early advection are only visible for high Rayleigh numbers, $\ra \geq  10^4$, and small time scales, $t <  10^{-3}$. 
For most microfluidic configurations investigating molecular solutes, these regimes might be difficult to observe even in a thick slit, as
$t = 10^{-3}$ in the above diagram does not exceed a few seconds for  $H=500~\mu$m and $D \geq 10^{-10}$~m$^2$/s.
On the other hand, both regimes of late advection and 1D dispersion should be easily observed  as the transition time between these two regimes $t_{\mathrm{III}\to \mathrm{IV}}$ is a few minutes for $D \simeq 10^{-10}$~m$^2$/s and $H=500~\mu$m.
In the case of colloidal dispersions, the regimes of early diffusion and early advection might  be observable as 
 much lower $D$ values lead to much longer time scales. For instance, $t = 10^{-3}$ is now a few minutes for  $H = 500~\mu$m and  $D= 10^{-12}$~m$^2$/s corresponding to colloids of  radius $100$~nm dispersed in water. 
%On the other hand, 
%le début du régime dispersif commence aux temps longs dans ce cas, 
%the dispersion regime becomes difficult to observe in this case, as the transition $t_{\mathrm{III}\to \mathrm{IV}}$  takes place around 40 minutes in this case.
%Note also that in most microfluidic configurations, the  regimes of early diffusion and early advection might be difficult to observe. 
%Indeed, even considering a molecular solute with a small diffusion coefficient, e.g.\, $D \simeq 10^{-10}$~m$^2$/s, in a relatively thick slit, $h=500~\mu$m, $t = 10^{-3}$ in the above diagram corresponds only to a few seconds.
%On the other hand, both the regime of late advection and Taylor dispersion should be easily observed, as the transition time between this two regimes $t_{\mathrm{III}\to \mathrm{IV}}$ is a few minutes with the values of the above numerical application.  

\subsection{Analogy with the case of a 2D porous layer\label{comparaisonJuanes}}
Szulczewski and Juanes~\cite{Szulczewski2013} studied in a different context (geological sequestration of CO$_2$ in an aquifer), a problem similar to the one  described in Fig.~\ref{fig1}, but considering a vertically confined porous layer of thickness $H$. Their theoretical model is also based on Eq.~(\ref{eq:convdim})  to describe the solute transport and Eq.~(\ref{eq:contdim}) for the overall mass conservation, but the pore velocity field $\mathbf{U}$  follows  Darcy's law given by:
\begin{eqnarray}
\mathbf{U} = -\frac{\kappa}{\rho_0 \nu \epsilon}\left[ \nabla P - (\rho(\Phi) - \rho_0)\mathbf{g} \right]\,, \label{eqDarcy}
\end{eqnarray}
where $\kappa$ is the permeability of the permeable rock and $\epsilon$ its porosity (see Sec.~\ref{sec:IIa} for the other notations).
Notice that Navier-Stokes equations Eq.~(\ref{eq:stokdim}) turns to Stokes equations when inertia is neglected, that does not reduce to Darcy's law
Eq.~(\ref{eqDarcy}). Indeed, Stokes equations include the diffusion of the momentum over the scale $H$ of the microfluidic channel, that brings a fundamental difference with  Darcy's law.

The characteritic velocity resulting from the Darcy's law  Eq.~(\ref{eqDarcy}) reads\begin{eqnarray}
U_D = \frac{\beta \Phi_i g \kappa}{\nu \epsilon}\,, \label{eqUDarcy}
\end{eqnarray}
and the corresponding Rayleigh number comparing diffusion and advection by the gravity current is:
\begin{eqnarray}
\widetilde{\ra} = \frac{\beta \Phi_i  g \kappa H}{\nu \epsilon D} \sim \frac{U_D H}{D}\,, \label{eqRatilideDarcy}
\end{eqnarray}
thus highlighting a scaling law with the  thickness different from the microfluidic case, $\widetilde{\ra} \propto H$ vs.\ $\ra \propto H^3$.
Interestingly, the range of Rayleigh numbers $\widetilde{\ra}$ involved in the context of CO$_2$ sequestration~\cite{Szulczewski2013} still corresponds to the range of  $\ra$ studied in the present work.

Despite the differences between these two models, Szulczewski and Juanes also reported five distinct   transport regimes in the porous rock that 
have strong similarities to those reported in Fig.~\ref{fig2}(a) for a microfluidic slit, see their names  in Table~\ref{tab:Szulczewski}.  
In particular, the concentration profiles depend on the same self-similar variables in each  regime, and the different transition times (displayed in the last column of 
Table~\ref{tab:transition} for our model) obey the same scaling laws with the Rayleigh number. 
We believe that these similarities are related to the linearity of the Darcy and Stokes equations in both problems.
Nevertheless, since the velocity fields are different in both configurations, we expect possibly different prefactors for the scaling laws. 
To confirm this point, we estimate the solute flux across the interface $z=0$ defined by: 
\begin{eqnarray}
f(t) = -\int_0^1 \left[\varphi u_z - \left(\frac{\partial \varphi}{\partial z}\right)\right]_{x,z=0}\text{d}x\,, \label{eq:definedF}
\end{eqnarray}
as  Szulczewski and Juanes also computed this quantity in each regime~\cite{Szulczewski2013}.
$f$ is estimated using Eqs.~(\ref{eq:solErf}) in Regimes~I and V, Eq.~(\ref{eq:uztheoini}) in Regime~II, following Ref.~\cite{Martin2011} in Regime~III, and using 
Eq.~(\ref{eq:diffeff}) in Regime~IV. As shown in Table~\ref{tab:Szulczewski}, $f$ for the microfluidic slit and $\tilde{f}$ for the porous layer show the same scaling laws
with $t$ and $\ra$ (resp. $\widetilde{\ra}$).
With the exception of the diffusion Regimes~I and~V, the large differences between the numerical prefactors (up to two orders of magnitude) confirm that both problems are fundamentally different. Relevant prefactors must be considered in potential comparisons with experimental works.

\begin{table}
\centering
\begin{tabular}{|l|l||l|l||}
  \hline
	\hline
		 \multicolumn{2}{|c||}{Microfluidic slit, this work }  & \multicolumn{2}{|c|}{2D porous layer~\cite{Szulczewski2013}} \\	
	  \hline
	\hline
  I- early diffusion&  $f \simeq \frac{1}{2(\pi t)^{1/2}} $ & I- early diffusion &  $\tilde{f} \simeq  \frac{1}{2(\pi t)^{1/2}}$ \\%& $t_{\mathrm{I}\to \mathrm{II}} \sim\ra^2$\\
	II- early advection&   $f \simeq 0.00345\, \ra$ & II- S-slumping & $\tilde{f} \simeq  0.186\, \widetilde{\ra}$ \\% $t_{\mathrm{II}\to \mathrm{III}}\sim \ra $\\
	III- late advection&   $f \simeq 0.0157\, \frac{\ra^{1/2}}{t^{1/2}}$ & III- straight-line slumping  & $\tilde{f}  \simeq 0.125 \,\frac{\widetilde{\text{Ra}}^{1/2}}{t^{1/2}}$ \\%& $t_{\mathrm{III}\to \mathrm{IV}} \sim 1$\\
	IV- 1D dispersion&  $f \simeq 0.00321\, \frac{\ra^{1/2}}{t^{3/4}}$ &   IV- Taylor slumping & $\tilde{f} \simeq 0.0238\, \frac{\widetilde{\ra}^{1/2}}{t^{3/4}}$ \\%&$t_{\mathrm{IV}\to \mathrm{V}} \sim \ra^2$ \\
	V- late diffusion&  $f \simeq \frac{1}{2(\pi t)^{1/2}} $ & V- late diffusion & $\tilde{f} \simeq  \frac{1}{2(\pi t)^{1/2}}$  \\% & ~\\
	\hline
  \hline
\end{tabular}
\caption{Solute flux defined by Eq.~(\ref{eq:definedF})  for a microfluidic slit and in a 2D porous layer~\cite{Szulczewski2013}.}
\label{tab:Szulczewski}
\end{table}

\section{The case of a 3D microfluidic channel \label{sec:3D}}

We now  consider the case of a microfluidic channel with a square cross-section, i.e. $\gamma = L/H = 1$ in Fig.~\ref{fig1}. Numerical simulations were performed for three Rayleigh numbers $\ra = 10^3$, $10^4$, and $10^5$ with a fixed Schmidt number $\sch = 10^3$.

\subsection{Transport regimes in the 3D case}

\begin{figure}[ht]
\begin{centering}
\includegraphics{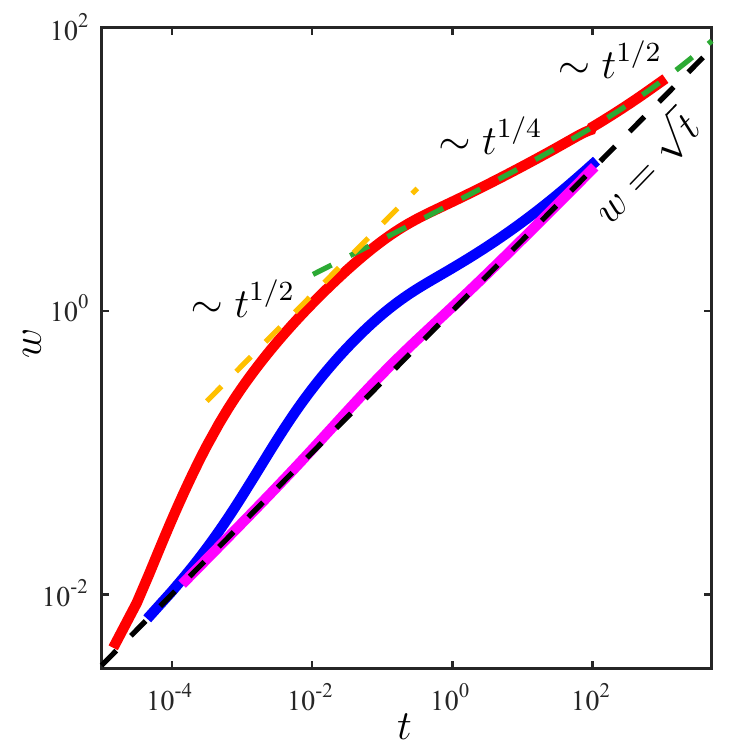}
\caption{Width of the mixing zone $w(t)$ obtained in the case of a channel with a square cross-section for $\sch = 10^3$ and $\ra = 10^5$ (red), $10^4$ (blue), $10^3$ (magenta).
%The thin red line corresponds to the slit case for  $\text{Ra}=10^5$ and $\text{Sc}=10^5$. 
Asymptotic models for $\ra=10^5$: 
the yellow dashed line given by Eq.~(\ref{eq:dispLockExchange3D}) corresponds to the regime of late advection; the green dashed line is computed from the numerical resolution of Eqs.~(\ref{eq:diffeff}) and~(\ref{eq:DeffTaylor}) with $\alpha \simeq 739872$. 
%(b) Width of the mixing zone $w$ vs.\ $t$ computed from the 3D model for $\sch = 10^3$ and $\ra = 10^5$ (red), $10^4$ (blue), $10^3$ (magenta). In (a) and (b), 
The black dashed line indicates the  diffusion law $w = \sqrt{t}$.  
  \label{FigWvsT_3D}}
\end{centering}
\end{figure}

Movie M2 supplied in the ESI shows the concentration and velocity fields in the planes $y=0$ and $z=0$ obtained from the numerical resolution of the 3D model for $\ra = 10^5$ and $\sch = 10^3$. This movie helps to identify the same succession of  regimes as  for the slit case. From the full numerical data, we computed again the width of the mixing zone $w(t)$ using Eqs.~(\ref{eq:defphi03D}) and (\ref{eq:defw}) for three different Rayleigh numbers $\ra = 10^3$, $10^4$ and $10^5$ and the same Schmidt number $\sch = 10^3$, see Fig.~\ref{FigWvsT_3D}.
%shows these data along with the curve $w(t)$  for the slit at $\ra = 10^5$ and $\sch = 10^5$ for comparison. 
The dispersion curves obtained  from the 3D model follow the same trends as the 2D ones shown in Fig.~\ref{fig2}(b).
These data lead to the same conclusion as for the slit case: buoyancy hardly affects the solute transport  at all time scales for $\ra \leq 10^3$
(the maximal value of the ratio $w/\sqrt{t}$ is only $\simeq 1.14$ at $t \simeq 0.1$ for $\ra = 10^3$). 
%and is slightly lower due to the increased viscous forces induced by the side walls. 

%%%%%%%%%%%%%%%%%%%%%%%%% 	
%%%%%%%%%%%%%%%%%%%%%%% 

These observations evidence that the description of the regimes presented above applies again for the 3D case.
Thereafter, we will not re-describe the regimes of early diffusion and early advection, in particular because they are  short or even hardly observable in most microfluidic experimental configurations. 
%coupled with the inertial transient  in the reported numerical dataset since $\sch = 10^3$. We also recall that these regimes correspond to very short times a priori difficult to observe experimentally. 
With respect to the late advection regime, the  analysis reported in Sec.~\ref{sec:lateadvection} can easily be adapted to the 3D case. The shape of the pseudo-interface separating the solution and the solvent is computed  using  Eq.~(\ref{eq:Martin}) with $f(\psi)$ and $\tilde{D}$ corresponding to a channel with a square cross-section, see Ref.~\cite{Martin2011} for details. We then calculated  the width $w(t)$ from the theoretical profiles, leading to:
\begin{eqnarray}
w \simeq 0.04104 \sqrt{\ra\,t}\,. \label{eq:dispLockExchange3D}
\end{eqnarray}
 This prediction fits well the data obtained from the 3D numerical simulation at $\ra = 10^5$ for time scales ranging from $t \simeq 3 \times 10^{-3}$ to $\simeq  3 \times 10^{-2}$, see Fig.~\ref{FigWvsT_3D}.
The numerical prefactor in Eq.~(\ref{eq:dispLockExchange3D}) ($0.04104$) is $16\%$ smaller than in Eq.~(\ref{eq:dispLockExchange}) for the slit case ($0.04879$), 
due to the increased viscous forces induced by the side walls in the 3D case. 

\begin{figure}[ht]
\begin{centering}
\includegraphics{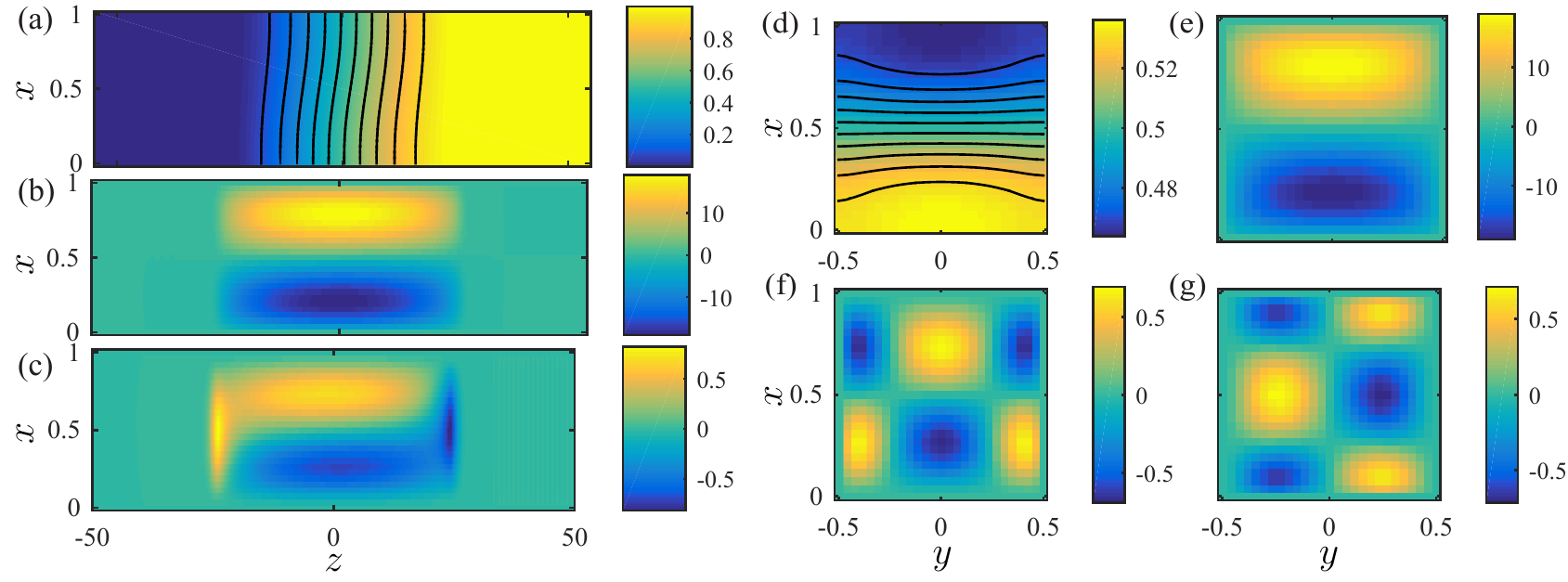}
\caption{3D case, $\ra = 10^5$, $\sch = 10^3$, and $t\simeq 4.7$. (a) Concentration field $\varphi(x,y=0,z,t)$. (b) Velocity field $u_z(x,y=0,z,t)$  and (c)
$u_x(x,y=0,z,t)$. (d)  $\varphi(x,y,z=0,t)$, (e) $u_z(x,y,z=0,t)$,  (f) $u_x(x,y,z=0,t)$, and (g) $u_y(x,y,z=0,t)$.
The black lines in  (a) and (d) are isoconcentration lines.
\label{FigTalorDisp3D}}
\end{centering}
\end{figure}

At later time scales $t \geq \mathcal{O}(1)$, Movie M2 evidences that diffusion almost homogenizes the concentration over the cross-section of the channel.
This is illustrated by Figs.~\ref{FigTalorDisp3D}(a) and~\ref{FigTalorDisp3D}(d) showing the concentration field in the planes $y=0$ and $z=0$ at 
$t \simeq 4.7$.
Figure~\ref{FigTalorDisp3D}, displaying also  the components of the velocity field in the same planes and at the same time, evidences a quasi-parallel flow along $z$
(due to the large extent $w$ of the longitudinal density gradient), and a secondary transverse flow as revealed by the components $u_x$ and $u_y$ in the $z=0$ plane. 
%%%%%%%%%%%%%%%%%%%%%%%%% 	
%%%%%%%%%%%%%%%%%%%%%%% 

All these results suggest, as for the 2D case of the slit, the existence of a dispersion regime  described by a 1D model, see Eq.~(\ref{eq:diffeff}).
However, the 3D case deserves particular attention because (i) to our knowledge, 
the expression  Eq.~(\ref{eq:DeffTaylor}) of the dispersion coefficient   cannot be found in the literature for a rectangular channel. It has been already calculated for a circular tube~\cite{Erdogan1967} and a slit \cite{Young1991}, and we derive it  for a rectangular cross-section below in Sec.~\ref{sec:TaylorDisp3D}; (ii) As mentioned by Chatwin and Ergogan~\cite{Erdogan1967} and evidenced in Fig.~\ref{FigTalorDisp3D}, a transverse flow exists in a 3D geometry, which might hinder the validity of the 1D dispersion model. This issue is addressed below in Sec.~\ref{ref:sectransverse}.

\subsection{1D dispersion regime in the 3D case \label{sec:TaylorDisp3D}}

%Movie MXX in the ESI helps to evidence that the shape of the velocity fields in the $z=0$ plane is almost invariant in this regime ($t \geq 1$).

The fact that $w \gg 1$  calls for the  "Taylor-like" approach leading to the 1D dispersion equation Eq.~(\ref{eq:diffeff}), see Appendix~\ref{app:effdisp2D} for the slit case.
In such a theoretical approach~\cite{Young1991},  transverse variations in concentration are assumed to be small:
\begin{eqnarray}
\varphi(x,y,z,t) = \varphi_0(z,t) + \varphi_1(x,y,z,t)\,,
\end{eqnarray}
with $\varphi_0$ given by Eq.~(\ref{eq:defphi03D}) and $\varphi_1 \ll \varphi_0$.
Averaging the transport equation Eq.~(\ref{eq:scconvadim}) over the cross-section of the channel leads to:
\begin{eqnarray}
\frac{\partial \varphi_0}{\partial t}  + \frac{\partial < u_z \varphi_1>}{\partial z}  = \frac{\partial^2 \varphi_0}{\partial z^2}\,, \label{eq:transpoAverage3D}
\end{eqnarray}
similarly to Eq.~(\ref{eq:transpoAverage}) for the case of a slit.
Subtracting this relation to Eq.~(\ref{eq:scconvadim}) gives:
\begin{eqnarray}
\frac{\partial \varphi_1}{\partial t}  + u_z \frac{\partial \varphi_0}{\partial z} + \mathbf{u}.\nabla \varphi_1 - \frac{\partial < u_z \varphi_1>}{\partial z}  =  \frac{\partial^2 \varphi_1}{\partial x^2}+\frac{\partial^2 \varphi_1}{\partial y^2}+\frac{\partial^2 \varphi_1}{\partial z^2}\,, \label{eq:fonda3D}
\end{eqnarray}
as Eq.~(\ref{eq:fonda2D}).
For $t \gg 1$ and $w \gg 1$, Eq.~(\ref{eq:fonda3D}) yields at leading order:
\begin{eqnarray}
u_z \frac{\partial \varphi_0}{\partial z} + u_y \frac{\partial \varphi_1}{\partial y} +  u_x \frac{\partial \varphi_1}{\partial x} \simeq	  \frac{\partial^2 \varphi_1}{\partial x^2}+\frac{\partial^2 \varphi_1}{\partial y^2}\,. \label{eq:fonda3D2}
\end{eqnarray}
In the 2D case of a slit, $u_y = 0$ and the continuity equation Eq.~(\ref{eq:sccontadim}) imposes  $u_x \sim u_z/w \ll u_z$. The term $u_x \partial_x \varphi_1$ in the above relation is thus negligible, and the derivation of the dispersion equation Eq.~(\ref{eq:diffeff}) is  straightforward, see Appendix~\ref{app:effdisp2D}.
 In the 3D case nevertheless, the continuity equation does not make it possible to  relate the scales of $u_y$ and $u_x$ to $u_z$, as there can exist (possibly large) secondary transverse flows verifying $\partial_x u_x + \partial_y u_y = 0$. %It is therefore not possible to neglect strictly the terms $ \mathbf{u}.\nabla \varphi_1$ in Eq.~(\ref{eq:fonda3D}) as for the slit case.
It is therefore a priori not possible to 
derive the dispersion equation Eq.~(\ref{eq:diffeff}) unless one neglects the terms $u_x \partial_x \varphi_1$ and $u_y \partial_y \varphi_1$  in Eq.~(\ref{eq:fonda3D}).
This point had been mentioned by Chatwin and Erdogan who studied the classical Taylor-Aris problem of the dispersion of a buoyant solute flowing in a circular tube~\cite{Erdogan1967}. 
They even showed that  the lateral mixing of the solute due to these secondary transverse flows could  lead to a decrease of the overall solute dispersion in a pressure-driven flow for some range of the Rayleigh number, see also Refs.~\cite{Barton1976,Smith1976}.
 
The data shown in Fig.~\ref{FigTalorDisp3D} along with Movie M2 in the ESI evidence that the maximal magnitude of these secondary flows is about $u_x \simeq u_y \simeq 1$ for $\ra = 10^5$, and we will assume 
as a first step that they do not significantly affect the  transport  of the solute. This assumption allows us to 
neglect  the terms $u_y \partial_y \varphi_1$ and $u_x \partial_x \varphi_1$ in Eq.~(\ref{eq:fonda3D}) as compared to the diffusive terms  $\partial_x^2 \varphi_1$ and 
$\partial_y^2 \varphi_1$, 
leading to:
\begin{eqnarray}
u_z \frac{\partial \varphi_0}{\partial z} \simeq  \frac{\partial^2 \varphi_1}{\partial x^2}+\frac{\partial^2 \varphi_1}{\partial y^2}\,, \label{eq:fonda3D_s}
\end{eqnarray}
similarly to Eq.~(\ref{eq:fonda2D_s}) for the slit. 

To compute the longitudinal flow $u_z$, we  first assume that it is described by the lubrication approximation because 
the extent of the mixing zone is large ($w \gg 1$):
\begin{eqnarray}
&&\frac{\partial^2 u_z}{\partial x^2}+\frac{\partial^2 u_z}{\partial y^2} =\frac{\partial p_\ell}{\partial z}\,, \label{eq:p00}\\
&&0 = \frac{\partial p_\ell}{\partial y} \,, \label{eq:p01}\\
&&0 = \frac{\partial p_\ell}{\partial x} + \ra\varphi_0\,,\label{eq:p02}
\end{eqnarray}
where $p_\ell$ is the pressure field associated to this longitudinal flow.
After integrating Eqs.~(\ref{eq:p01}) and~(\ref{eq:p02}) and inserting the resulting pressure field in Eq.~(\ref{eq:p00}), one finds:
\begin{eqnarray}
\frac{\partial^2 u_z}{\partial x^2}+\frac{\partial^2 u_z}{\partial y^2} = \ra \frac{\partial \varphi_0}{\partial z} \left(\frac{1}{2}-x\right) \,, \label{eq:fonda3D_uz}
\end{eqnarray}
the term $1/2$ ensuring that $<u_z> = 0$.

From Eqs.~(\ref{eq:fonda3D_s}) and~(\ref{eq:fonda3D_uz}), it is therefore possible to calculate both $u_z$ and $\varphi_1$, and finally to compute the dispersion term $<u_z \varphi_1>$ in Eq.~(\ref{eq:transpoAverage3D}) leading ultimately to the same dispersion equation Eq.~(\ref{eq:diffeff}) but with
a different $\alpha$ value for $D_\mathrm{eff}$ in Eq.~(\ref{eq:DeffTaylor}).
Appendix~\ref{app:effdispDD} reports solutions of Eqs.~(\ref{eq:fonda3D_s}) and~(\ref{eq:fonda3D_uz}) using Fourier series for a rectangular channel of arbitrary aspect ratio $\gamma$,
see Eqs.~(\ref{eq:uzFourier}) and~(\ref{eq:varphi12D}). These calculations allow us to compute the numerical values of $\alpha$ as a function of $\gamma$, see Fig.~\ref{PrefAlpha} in Appendix~\ref{app:effdispDD}.
For a square cross-section, $\alpha \simeq 739872$, and 
$\alpha$ tends towards the value derived for the slit $\alpha = 362880$ for infinitely thin channel,  unlike the Taylor-Aris coefficient in a  rectangular channel~\cite{Chatwin1982,Ajdari:06}. 

\begin{figure}[ht]
\begin{centering}
\includegraphics{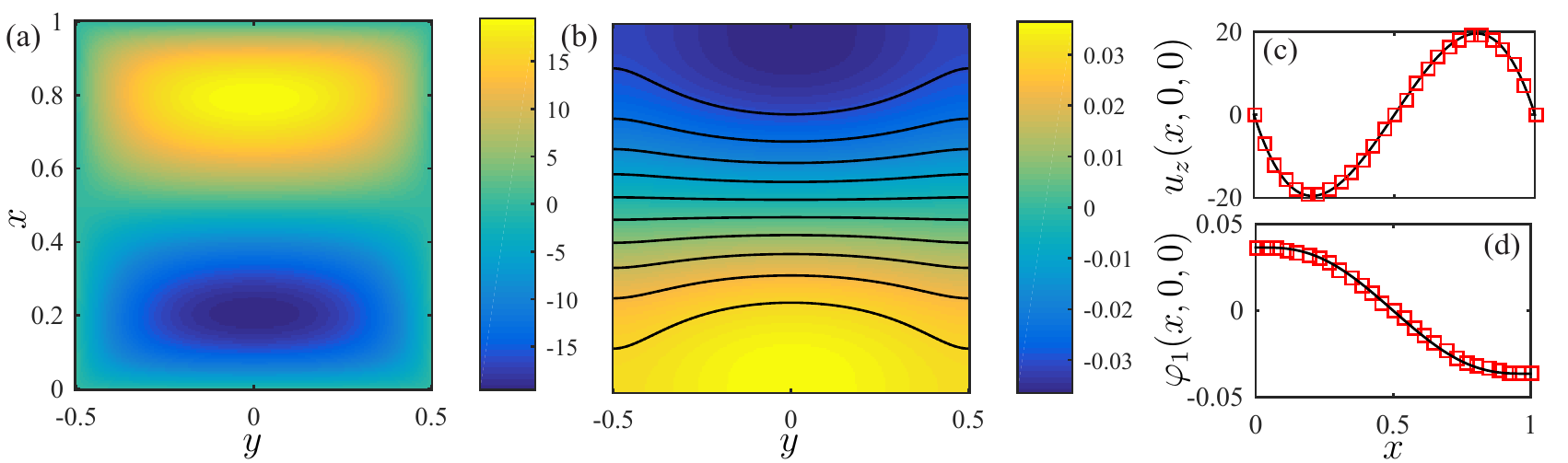}
\caption{(a) $u_z(x,y,z=0)$ and (b) $\varphi_1(x,y,z=0)$ calculated using   Eqs.~(\ref{eq:uzFourier}) and~(\ref{eq:varphi12D})  at $t \simeq 4.7$, see the corresponding numerical data shown in Fig.~\ref{FigTalorDisp3D}(d) and (e).  
The longitudinal density gradient is calculated by the numerical resolution of the dispersion equation Eq.~(\ref{eq:diffeff}).
(c) and (d) show the comparisons along the line $y=z=0$ between the numerical data reported in Fig.~\ref{FigTalorDisp3D}(d) and (e) (squares) and the theoretical profiles given by
Eqs.~(\ref{eq:uzFourier}) and~(\ref{eq:varphi12D}) 
 (black lines). 
\label{FigTalorDisp3DCompar}}
\end{centering}
\end{figure}

Figures~\ref{FigTalorDisp3DCompar}(a) and~\ref{FigTalorDisp3DCompar}(b)
shows both the component $u_z$ and $\varphi_1$ in the plane $z=0$ calculated using Eqs.~(\ref{eq:uzFourier}) and~(\ref{eq:varphi12D}) and the 1D dispersion model Eq.~(\ref{eq:diffeff}) at $t \simeq 4.7$.
These data well match the numerical data reported in Fig.~\ref{FigTalorDisp3D}(d) and (e) at the same time, see also the comparisons along the line $y=z=0$ shown in Figs.~\ref{FigTalorDisp3DCompar}(c) and (d).
Consequently, the section-averaged concentration profiles $\varphi_0(z,t)$ are correctly predicted by the  1D  model (data not shown). These profiles make it possible to compute the width of the mixing zone $w(t)$ [Eq.~(\ref{eq:defw})] plotted in Fig.~\ref{FigWvsT_3D} along with the data obtained from the 3D numerical simulation for $\ra=10^5$. The 1D dispersion approach accounts well for the  spreading  of the solute for $t \geq 0.1$.  Moreover, Eq.~(\ref{eq:dispTA}) giving the $w \sim t^{1/4}$ behavior, allows us to compare the 2D and 3D configurations in this regime. 
Eq.~(\ref{eq:dispTA}) indeed indicates that the ratio $w_{\text{2D}}(t)/w_{\text{3D}}(t)$ between the 2D and the 3D cases is  $(\alpha_{\text{2D}} /\alpha_{\text{3D}})^{1/4} \simeq (362880/739872)^{1/4} \simeq 0.84 \leq 1$,   again due to the side walls increasing viscous forces in the 3D case.

%%%%%%%%%%%%%%%%%%%%%%%%% 	
%%%%%%%%%%%%%%%%%%%%

\subsection{Transverse flows and validity range of the 1D dispersion equation \label{ref:sectransverse}}

The comparisons shown in Sec.~\ref{sec:TaylorDisp3D} evidence the validity of the 1D  dispersion equation to describe the overall solute transport  
at least for $\ra \lesssim 10^5$, although the transverse flow is not considered in the  1D   model. In the following, we derive the expression of the transverse flow with the assumption that it does not significantly affect the concentration field. Then, we determine the critical Rayleigh number for which this hypothesis no longer holds.

Our calculations make it possible to compute the transverse variations in concentration $\varphi_1$, see  Fig.~\ref{FigTalorDisp3DCompar}(b). These variations evidence density gradients along $y$ due to the presence of the lateral walls which impact the longitudinal flow, and thus the solute distribution.
These density gradients are responsible for a secondary flow  because they cause a pressure gradient along $y$. 
We compute this flow assuming that it is locally invariant along $z$ because of the large extent of the mixing zone ($w\gg 1$) and solenoidal, i.e.\ 
$\partial_y u_y + \partial_x u_x = 0$, to ensure the global mass conservation.
%We will therefore from the Navier-Stokes equation Eq.~(\ref{eq:scstokadim}), still assuming that they do not impact the transport and thus the concentration field $\varphi_1$ that generates them.
This flow is thus solution of:
 \begin{eqnarray}
&& 0 = \frac{\partial p_t}{\partial z}\,,\label{eq:trans30}\\
&&\frac{\partial^2 u_y}{\partial x^2}+\frac{\partial^2 u_y}{\partial y^2} =  \frac{\partial p_t}{\partial y}\,, \label{eq:trans3a}\\ 
&&\frac{\partial^2 u_x}{\partial x^2}+\frac{\partial^2 u_x}{\partial y^2} = \frac{\partial p_t}{\partial x} + \ra\varphi_1\,, \label{eq:trans3b}
\end{eqnarray}
where  $p_t$ is the pressure field associated to this 2D solenoidal flow.
As in the work of Chatwin and Erdogan~\cite{Erdogan1967},  the global flow is therefore assumed to be the superposition of the longitudinal flow described by Eqs.~(\ref{eq:p00}-\ref{eq:p02}) with the secondary transverse flow determined by  Eqs.~(\ref{eq:trans30}-\ref{eq:trans3b}). 

%%%%%%%%%%%%%%%%%%%%%%%%% 	
\begin{figure}[ht]
\begin{centering}
\includegraphics{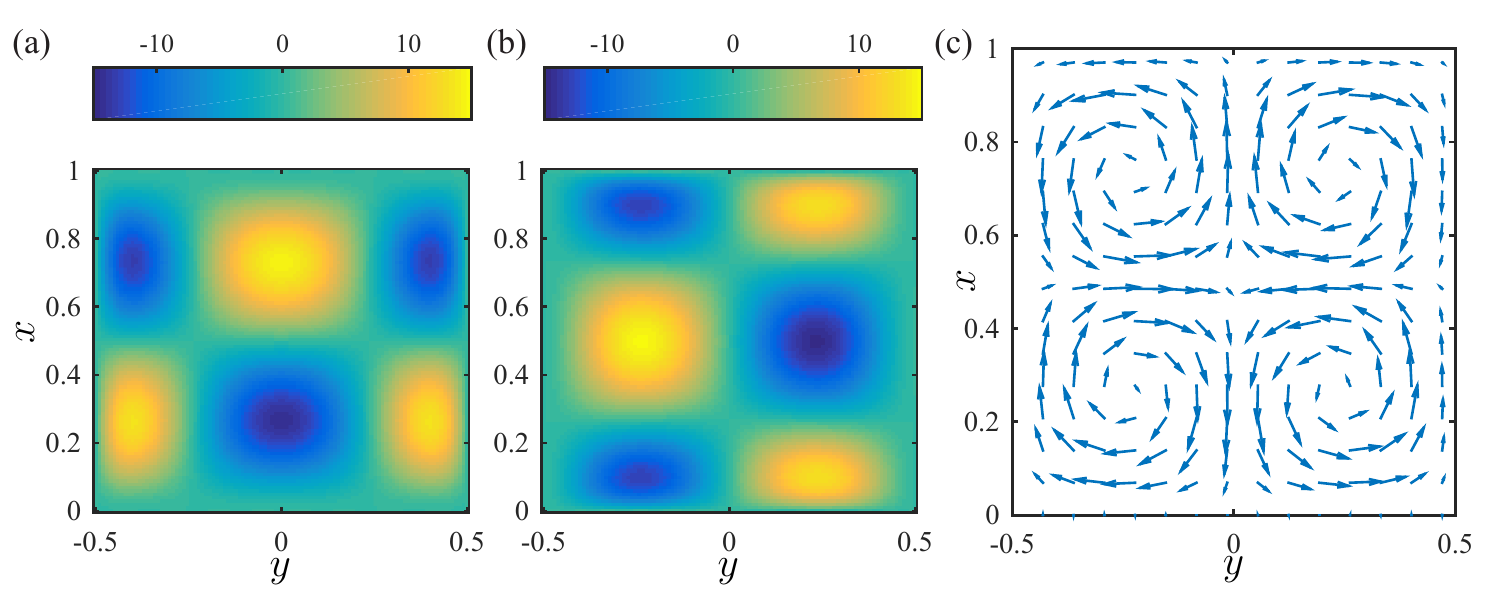}
\caption{Components $u_x$ (a) and $u_y$ (b) scaled by $10^{-8} \, \left(\ra \frac{\partial \varphi_0}{\partial z}\right)^2$ computed from the resolution of the biharmonic equation 
 Eq~(\ref{eq:FondaTrans}). (c) Corresponding transverse flow field in the plane $x$-$y$. This solenoidal flow field results from the density gradients given by $\varphi_1$, see Eq.~(\ref{eq:FondaTrans}). 
\label{Fig:FigTransverse}}
\end{centering}
\end{figure}
%%%%%%%%%%%%%%%%%%%%%%%  

An analytical approximation of the solution of Eqs.~(\ref{eq:trans30}-\ref{eq:trans3b}) is computed in Appendix~\ref{app:Transverse} for a rectangular cross-section of arbitrary aspect ratio $\gamma$ using a method described by Shankar {\it et al.} ~\cite{Shankar1993,Shankar2002}. 
Figure~\ref{Fig:FigTransverse} shows the theoretical prediction of this secondary transverse flow for $\gamma=1$.
The vector velocity field shown in Fig.~\ref{Fig:FigTransverse}(c) simply corresponds to the solenoidal flow associated to the density gradients  revealed in Fig.~\ref{FigTalorDisp3DCompar}(b).
The predicted flow pattern compares well with the numerical data reported in Fig.~\ref{FigTalorDisp3D}(f) and (g), and 
our calculations show in particular that the maximal values of the components $u_x$ and $u_y$ are:
\begin{eqnarray}
\mathrm{max}(u_x) \simeq \mathrm{max}(u_y) \simeq 14 \times 10^{-8} \left(\ra \frac{\partial \varphi_0}{\partial z}\right)^2\,.  \label{eq:uxuymax}
\end{eqnarray}

\begin{figure}[ht] 
\begin{centering}
\includegraphics{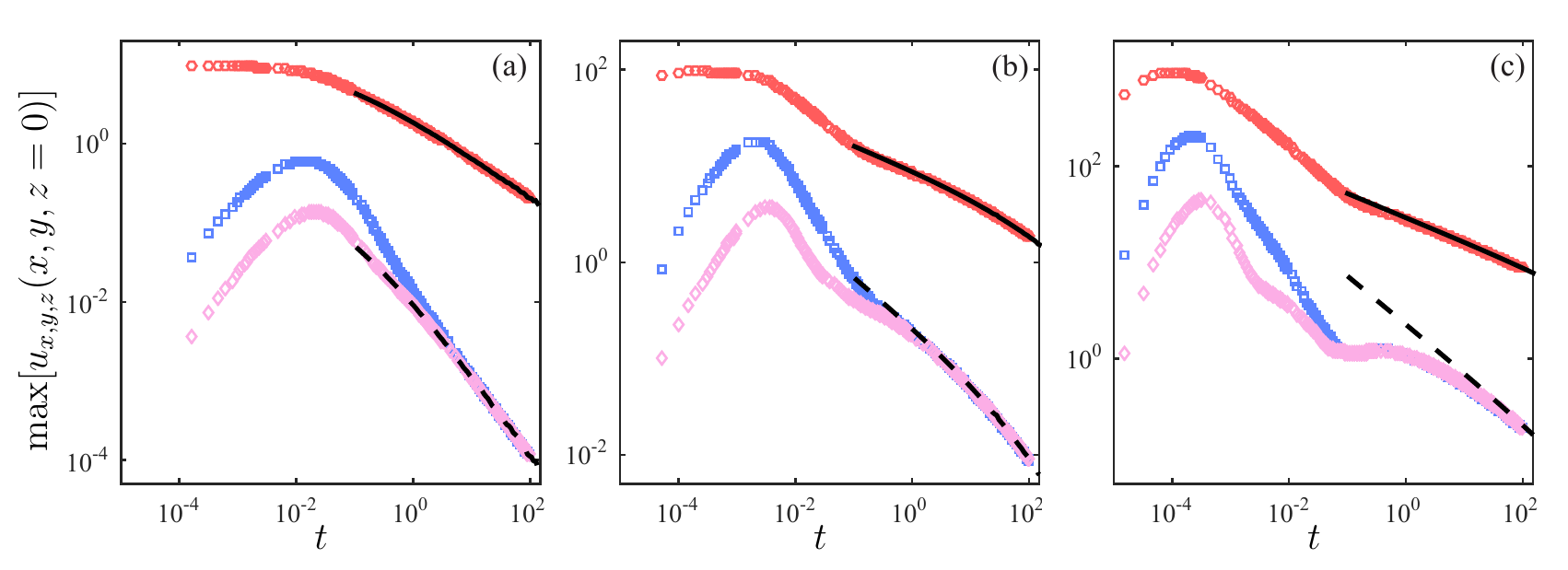}
\caption{Maximal value of the components $u_z$ (red), $u_x$ (blue) and  $u_y$ (magenta) in the plane $z=0$ computed from the numerical 3D model at (a) $\ra =  10^3$, (b)
$10^4$, and (c) $10^5$ ($\sch= 10^3$ for these three cases). The black lines are the prediction given by Eq.~(\ref{eq:uzFourier}), and  the dotted line the prediction given by Eq.~(\ref{eq:uxuymax}).
In these theoretical predictions, the longitudinal density gradient $\partial_z \varphi_0$ at $z=0$ is calculated from the numerical solution of the 1D dispersion Eq.~(\ref{eq:diffeff}).
\label{FigComparTransverse3D}}
\end{centering}
\end{figure}

To go a step further into the comparison,  Fig.~\ref{FigComparTransverse3D} displays the maximal values of the components $u_x$, $u_y$ and $u_z$ in the transverse plane 
$z=0$ vs.\ time $t$ obtained from the 3D model for three Rayleigh numbers, $\ra = 10^3$, $10^4$, and $10^5$. 
%%%%%%%%%%%%%%%%%%%%%%%%% 	
%%%%%%%%%%%%%%%%%%%%
The theoretical prediction given by Eq.~(\ref{eq:uzFourier}) for the longitudinal flow $u_z$ correctly fits the numerical data even at $\ra = 10^5$ for $t \geq 1$.
The theoretical transverse components given by Eq.~(\ref{eq:uxuymax}) accounts well for the data at $\ra = 10^3$ and $\ra = 10^4$ [with the  gradient of $\varphi_0$ at $z=0$ computed from the numerical solution of Eq.~(\ref{eq:diffeff})], but significant discrepancies are observed for $\ra = 10^5$. For this Rayleigh number,  $u_x \simeq u_y \simeq 1$ when entering the  1D dispersion regime ($t \geq 1$), and advection by the secondary transverse flow is no more negligible compared to the transverse diffusion. 
It is remarkable to see that the theoretical prediction given by Eq.~(\ref{eq:uxuymax}) makes it possible to  predict this transition quantitatively.
More precisely, the concentration profiles in the dispersion regime are correctly described by the asymptotic approximation Eq.~(\ref{eq:appAlb}) when buoyancy dominates,
and one can  thus compute the longitudinal gradient at $z=0$ to write Eq.~(\ref{eq:uxuymax}) as follows:
\begin{eqnarray}
\mathrm{max}(u_x) \simeq \mathrm{max}(u_y) \simeq 2.2 \times 10^{-5}\,\frac{\ra}{\sqrt{t}}\,,  \label{eq:uxuymaxvst}
\end{eqnarray}
for a square cross-section ($\alpha \simeq 739872$).
Therefore, the impact of transverse flows on the solute transport is expected to be negligible for  $\mathrm{max}(u_x) \simeq \mathrm{max}(u_y) \leq 1$ leading to the criterion:
\begin{eqnarray}
t \geq 4.9 \times 10^{-10}\,\ra^2 \,,  \label{eq:uxuymaxvstc}
\end{eqnarray}
about $t \geq 4.9$ for $\ra = 10^5$, in a remarkable agreement with the data of Fig.~\ref{FigComparTransverse3D}(c) showing a discrepancy between the numerical solutions ($u_x$,$u_y$) and the predictions  given by Eq.~(\ref{eq:uxuymax}) for $t \leq 10$. 
Nevertheless, the transverse flow does not yet significantly change the longitudinal dispersion even for $\ra = 10^5$ at $t > 1$ (see the good agreement between the 1D dispersion model and the 3D numerical simulation noted in Sec.~\ref{sec:TaylorDisp3D}).
However, our theoretical predictions suggest that a significant impact on the overall solute dispersion is expected for higher Rayleigh numbers. The secondary transverse flow probably leads in this case to  a  decrease of the overall solute dispersion because it contributes to {\it mix} the solute laterally, as also shown by Chatwin and Erdogan in a different context, the 
Taylor-Aris dispersion of a buoyant solute in a pressure-driven flow~\cite{Erdogan1967,Barton1976,Smith1976}. In most microfluidic experimental configurations,  these  effects can a priori be neglected because $\ra \leq 10^5$.

\section{Conclusions}
In the present work, we have studied in detail the impact of buoyancy on solute spreading in two distinct microfluidic geometries: a 2D slit and a microchannel with a square cross-section, in particular through analytical predictions fully validated by precise numerical resolutions of the transport equations. 
One of the main results of our study is to show that for $\ra \leq 10^3$, solutal free convection does not impact  solute diffusion at all time scales.
Beyond this result, our theoretical predictions give also for larger Rayleigh numbers, the time scales (or density gradients) for which buoyancy no longer impacts molecular diffusion, see the diagram in Fig.~\ref{FigDiagramme}. 
%Beyond this result, our theoretical predictions also make it possible to know at which time scales (and thus for which density gradients), these gravity currents no longer impact molecular diffusion for larger Rayleigh numbers, see in particular the diagram in Fig.~\ref{FigDiagramme}.
Moreover, these same theoretical predictions allow to estimate analytically the gravity currents, whatever their role on solute transport.
It is worth remembering that these flows may impact the transport of other species dispersed in the flow, even though they do not affect the gradients of concentration of the {\it active} species that generate them.  As an example, for the experimental case mentioned in Introduction, interdiffusion between water and a NaCl aqueous solution at 1~M in a microfluidic slit of height $H=100~\mu$m,  free convection is not expected to impact the diffusive mixing because $\ra \simeq 230$. Nevertheless, the typical
longitudinal velocity  defined by Eq.~(\ref{eq:defuz}) is about $\bar{u}_z \simeq 25~\mu$m/s in the early regimes of diffusion and advection, and still  $\bar{u}_z \simeq 5~\mu$m/s  for the time scale $T = H^2/D$ (see Table~I), and could  significantly advect  less mobile species dispersed in the solutions. 
The precise control of transport conditions in microfluidic geometries thus possibly opens the way to flow control induced by solute gradients.

For the 3D case of a rectangular cross-section channel, our work brings for the first time (to our knowledge) estimates of the  1D dispersion coefficient 
describing the transport of the solute  at long time scales for any aspect ratio $\gamma= L/H$.  Our work also highlights  a subtle point related to 3D geometries: 
the order of magnitude of the transverse flows cannot be determined from  
the lubrication approximation alone. In the case studied here, these flows, induced by transverse density gradients, remain moderate up to $\ra = 10^5$, and the overall solute spreading is correctly described by a 1D dispersion equation. 
However, our work predicts that these flows could play a role at higher  Rayleigh numbers  for experimental situations outside the field of application of microfluidics.

It could also be relevant to study more in details the case of shallow channels commonly encountered in microfluidic applications, and in particular to study more finely the transition between a channel with a large aspect ratio, $\gamma \gg 1$, and the slit.
Indeed, the case $\gamma \gg 1$ deserves more attention because a new time scale appears, the diffusion time over the width of the channel  $\sim \gamma^2$, see Ref.~\cite{Ajdari:06} investigating this issue for the case of the Taylor-Aris dispersion.

Finally, we considered in our work the case of an ideal binary solution in the framework of the Boussinesq approximation, see Eqs.~(\ref{eq:stokdim}--\ref{eq:convdim}). Microfluidic technologies allow a very fine control of the transport conditions (especially mass and momentum), and thus to study  interdiffusion in more complex mixtures. It would then be useful to go beyond the model described by Eqs.~(\ref{eq:stokdim}--\ref{eq:convdim}) to include this complexity: change in viscosity and diffusion coefficient as a function of concentration, role of the reference frame (volume velocity / mass velocity)~\cite{Joseph:96,Brenner:05}, 
other transport mechanisms (e.g. diffusio-osmosis), etc.
In this context, we hope that our work will make it possible to disentangle the role played  by solutal free convection from other transport phenomena.

\section*{Acknowledgements}
This work was performed using HPC resources from the "M\'esocentre" computing center of CentraleSup\'elec and \'Ecole Normale Sup\'erieure Paris-Saclay supported by CNRS and R\'egion \^Ile-de-France (http://mesocentre.centralesupelec.fr/).
JBS also thanks Y. Hallez for discussions concerning the lock-exchange problem and ANR OSMOCHIP (ANR-18-CE06-0021) as well as Solvay and CNRS for
funding.

\appendix
\section{Buoyancy-driven flow at early stage for the case of a slit \label{app:flowearlystage}}
After a transient corresponding to the diffusion of the momentum across the slit, the velocity field is solution of the Stokes equation Eq.~(\ref{eq:scstokadimStokes}).
Introducing the stream function $\psi(x,z)$ defined by:
\begin{eqnarray}
&&u_x = \frac{\partial \psi}{\partial z}~~\mathrm{and}~~u_z = -\frac{\partial \psi}{\partial x}\,, \label{eq:defstreamslit}
\end{eqnarray}
Eq.~(\ref{eq:scstokadimStokes}) is equivalent to:
\begin{eqnarray}
\Delta^2 \psi = \ra \delta(z)\,, \label{eq:biharmslit}
\end{eqnarray}
with $\delta(z)$ the Dirac function and $\Delta^2$ the biharmonic operator in 2D.
The no-slip boundary conditions impose:
\begin{eqnarray}
\frac{\partial \psi}{\partial z}(x=0~\mathrm{and}~1,z)=0~~\mathrm{and}~~\frac{\partial \psi}{\partial x}(x=0~\mathrm{and}~1,z)=0\,.
\end{eqnarray}
The flow is expected to vanish far from $z=0$ and we impose $\psi(x,z\to \pm \infty) = 0$. 
Because $\text{d}\psi = -u_z \text{d}x + u_x \text{d}z$, integrating $\text{d}\psi$ over a contour along $x=0$ or $x=1$ starting from $z \to \pm \infty$ imposes:
\begin{eqnarray}
\psi(x=0~\mathrm{and}~1,z)=0\,.
\end{eqnarray}
We define the Fourier transform of $\psi(x,z)$ by:
\begin{eqnarray}
\tilde{\psi}(x,k) = \frac{1}{\sqrt{2 \pi }}\int_{-\infty }^{\infty}{\rm d} z \psi(x,z) e^{i k z}\,.
\end{eqnarray} 
Eq.~(\ref{eq:biharmslit}) turns to:
\begin{eqnarray}
\frac{\partial^4 \tilde{\psi}}{\partial x^4}-2 k^2 \frac{\partial^2 \tilde{\psi}}{\partial x^2} + k^4 \tilde{\psi} = \frac{\ra}{\sqrt{2\pi}}\,.
\end{eqnarray}
The solution of the ordinary differential equation is:
\begin{eqnarray}
&&\tilde{\psi}(x,k) = \ra \frac{k + k (x-1) \cosh(k x) - k x \cosh(k - k x) + \sinh(k) - 
   \sinh(k x) - \sinh(k - k x)}{\sqrt{2 \pi }\, k^4 (k+\sinh (k))},
\end{eqnarray}
for the above boundary conditions.
The stream function is then computed from the inverse Fourier transform of $ \tilde{\psi}(x,k)$
%\begin{eqnarray}
%\psi(x,z) = \frac{1}{\sqrt{2 \pi }}\int_{-\infty }^{\infty}{\rm d}k \tilde{\psi}(x,k) e^{-i k z}
%\end{eqnarray}
and the velocity field is found using Eqs.~(\ref{eq:defstreamslit}). These calculations lead in particular to Eqs.~(\ref{eq:uxtheoini}) and~(\ref{eq:uztheoini}) given in Sec.~\ref{sec:earlydiffearlyadv}. 
Due to the symmetry of the equations along the  $x=1/2$ plane, the maximum of $u_x$ occurs at $x=1/2$ for all $z$ values.

%%%%%%%%%%%%%%%%%%%%
\section{Derivation of the advection-dispersion equation for the case of a slit \label{app:effdisp2D}}
We assume:
\begin{eqnarray}
\varphi(x,z,t) = \varphi_0(z,t) + \varphi_1(x,z,t)\,,
\end{eqnarray}
with $\varphi_0$ given by Eq.~(\ref{eq:defphi03D}) and $ \varphi_1  \ll \varphi_0$.
Averaging the transport equation Eq.~(\ref{eq:scconvadim}) over the height of the slit with the help of the continuity relation Eq.~(\ref{eq:sccontadim}), leads to:
\begin{eqnarray}
\frac{\partial \varphi_0}{\partial t}  + \frac{\partial < u_z \varphi_1>}{\partial z}  = \frac{\partial^2 \varphi_0}{\partial z^2}\,. \label{eq:transpoAverage}
\end{eqnarray}
Subtracting this relation to  Eq.~(\ref{eq:scconvadim}) yields:
\begin{eqnarray}
\frac{\partial \varphi_1}{\partial t}  + u_z \frac{\partial \varphi_0}{\partial z} + \mathbf{u}.\nabla \varphi_1 - \frac{\partial < u_z \varphi_1>}{\partial z}  =  \frac{\partial^2 \varphi_1}{\partial x^2}+\frac{\partial^2 \varphi_1}{\partial z^2}\,. \label{eq:fonda2D}
\end{eqnarray}
If we now assume that the scale $w$ of the gradient along $z$ verifies $w \gg 1$, the continuity equation imposes $u_x \sim  u_z/w$, and Eq.~(\ref{eq:fonda2D}) reduces to:
\begin{eqnarray}
u_z \frac{\partial \varphi_0}{\partial z} \simeq  \frac{\partial^2 \varphi_1}{\partial x^2}\,, \label{eq:fonda2D_s}
\end{eqnarray}
provided that $t \gg 1$ and $\varphi_1 \ll \varphi_0$.
Similarly, by neglecting the inertial term in Eq.~(\ref{eq:scstokadim}) and assuming again $w \gg 1$, one find Eq.~(\ref{eq:compuzTaylor}) for the component $u_z$, 
and $u_x$ given by the continuity equation Eq.~(\ref{eq:sccontadim}).
Integration of Eq.~(\ref{eq:fonda2D_s}) along with the impermeability boundary condition and $<\varphi_1>=0$  gives:
 \begin{eqnarray}
\varphi_1 = -\frac{\text{Ra}}{1440} \left(\frac{\partial  \varphi_0}{\partial  z}\right)^2 \left( 12x^5- 30 x^4 + 20x^3- 1\right)\,.  \label{eq:varphi1} 
\end{eqnarray}
Inserting Eq.~(\ref{eq:varphi1}) into Eq.~(\ref{eq:transpoAverage}) leads to the dispersion equation Eq.~(\ref{eq:diffeff}) 
with the dispersion coefficient given by Eq.~(\ref{eq:DeffTaylor}).
Comparison of the solution of Eq.~(\ref{eq:diffeff}) with the data obtained from the full 2D model shows that the 1D dispersion model is actually valid as soon as $t \geq 1$.

%%%%%%%%%%%%%%%%%%%%
\section{Derivation of the advection-dispersion equation for the case of a microfluidic channel with a rectangular cross-section \label{app:effdispDD}}

The solution of Eq.~(\ref{eq:fonda3D_uz}) is found by a Fourier sum noticing first that:
\begin{eqnarray}
&&\frac{1}{2}-x = \sum_{n=2, \mathrm{even}}^{\infty}\frac{2}{n\pi} \sin(n\pi x)\,~~\text{for}~~ x \in ]0-1[\,.
\end{eqnarray}
The Fourier series representing the solution of Eq.~(\ref{eq:fonda3D_uz}) along with the no-slip boundary conditions at the solid walls is then:
\begin{eqnarray}
&&u_z =  \ra \frac{\partial \varphi_0}{\partial z} \sum_{n=2,\mathrm{even}}^{\infty}  \frac{2}{(n\pi)^3} \left(\frac{\cosh(\pi n y)}{\cosh(\pi n \gamma/2)} -1 \right)  \sin(n\pi x)\,,    \label{eq:uzFourier}
\end{eqnarray}
with $\gamma = L/H$, see Ref.~\cite{Bruus} for a similar calculation of the pressure-driven velocity profile in a rectangular channel.
The Fourier series representing $\varphi_1$ can now be found using Eq.~(\ref{eq:fonda3D_s}) along with the impermeability boundary conditions at the solid walls and the constraint $< \varphi_1>= 0$, leading after calculations to:
\begin{eqnarray}
\varphi_1 = \frac{8\ra}{\pi^6} \left(\frac{\partial \varphi_0}{\partial z}\right)^2  \sum_{n=1,\mathrm{odd}}^{\infty} \sum_{p=2,\mathrm{even}}^\infty \frac{\cos(n\pi x) }{p^2(p^2-n^2)^2}
&&\bigg[ \frac{\cosh(\pi p y)}{\cosh(\pi p \gamma/2)} \notag \\
&& - \frac{p}{n} \tanh(\pi p\frac{\gamma}{2}) \frac{\cosh(\pi n y)}{\sinh(\pi n \gamma/2)}+\frac{p^2-n^2}{n^2} \bigg]. \label{eq:varphi12D}
\end{eqnarray}
   The dispersion term $<u_z \varphi_1>$ in Eq.~(\ref{eq:transpoAverage3D}) can now be evaluated leading finally to the dispersion equation Eq.~(\ref{eq:diffeff}) with 
$D_\mathrm{eff}$ given by Eq.~(\ref{eq:DeffTaylor}). 
The numerical prefactor $\alpha$ is given by:
\begin{eqnarray}
&&\frac{1}{\alpha} =  \frac{1}{362880}-\frac{64}{\gamma \pi^{11}} \sum_{n=1,\mathrm{odd}}^{\infty} \sum_{p=2,\mathrm{even}}^\infty \sum_{q=2,\mathrm{even}}^{\infty}\frac{1}{n^2 p^2 q^2 (p^2-n^2)^2(q^2-n^2)}\bigg[ \frac{n^2 \left(p \tanh \left(\frac{\pi  \gamma  p}{2}\right)-q \tanh \left(\frac{\pi  \gamma  q}{2}\right)\right)}{p^2-q^2}\notag\\
&&-\frac{n p \tanh \left(\frac{\pi \gamma  p}{2}\right) \left(n-q \coth \left(\frac{\pi  \gamma  n}{2}\right) \tanh \left(\frac{\pi  \gamma  q}{2}\right)\right)}{n^2-q^2}+\frac{(p^2-n^2) \tanh \left(\frac{\pi  \gamma  q}{2}\right)}{q}+ \frac{(p^2-n^2) \tanh \left(\frac{\pi  \gamma  p}{2}\right)}{p} \bigg]\,, \label{eq:prefalpha}
\end{eqnarray} 
and therefore only depends on the aspect ratio $\gamma$ of the channel.

\begin{figure}[ht]
\begin{centering}
\includegraphics{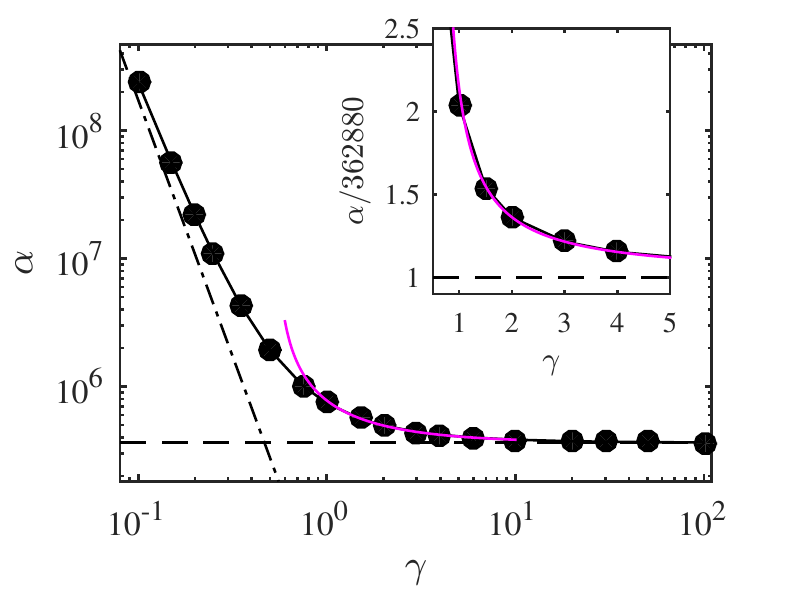}
\caption{Prefactor $\alpha$ in Eq.~(\ref{eq:DeffTaylor}) for a rectangular cross-section of aspect ratio $\gamma$. The dashed dotted line is the asymptotic bahavior Eq.~(\ref{eq:alphaasympt1}) valid for $\gamma \to 0$.
The magenta line is the approximation Eq.~(\ref{eq:alphaasympt2}) valid for $\gamma \gg 1$. The dashed line is the value for a slit $\alpha = 362880$.
Inset: zoom on the $\gamma$ range $1$--$4$.
\label{PrefAlpha}}
\end{centering}
\end{figure}

Asymptotic approximations of $\alpha$ can be found for $\gamma \to 0$:
\begin{eqnarray}
\frac{1}{\alpha} = \frac{\gamma ^4}{17280} + \mathcal{O}(\gamma^6)\,. \label{eq:alphaasympt1}
\end{eqnarray}
For a wide slit, i.e.\ $\gamma \gg 1$, the terms $\tanh$ and $\coth$ in Eq.~(\ref{eq:prefalpha}) are close to $\simeq 1$, and one thus find:
\begin{eqnarray}
&&\frac{1}{\alpha} \simeq  \frac{1}{362880}-\frac{1}{681432\gamma}\,,  \label{eq:alphaasympt2}
\end{eqnarray} 
where the factor $\simeq 681432$ is found by calculating numerically the sum in Eq.~(\ref{eq:prefalpha}).
Figure~\ref{PrefAlpha} displays the values of $\alpha$ calculated using Eq.~(\ref{eq:prefalpha}) for several aspect ratios $\gamma$ along with the 
asymptotic behaviors Eqs.~(\ref{eq:alphaasympt1}) and (\ref{eq:alphaasympt2}).
%%%%%%%%%%%%%%%%%%%%%%%%% 	
%%%%%%%%%%%%%%%%%%%%%%% 	
It should be noted that the  approximation Eq.~(\ref{eq:alphaasympt2}) valid for thin channels $\gamma \gg 1$ yields correct estimates of $\alpha$ within $<0.5 \%$ for $\gamma \geq 1.5$.
The value for a square cross-section is $\alpha \simeq 739872$.

%%%%%%%%%%%%%%%%%%%%
\section{Derivation of the secondary transverse flow\label{app:Transverse}}

We introduce the  stream function $\psi(x,y)$ defined by:
\begin{eqnarray}
&&u_x = \frac{\partial \psi}{\partial y}~~\mathrm{and}~~u_y = -\frac{\partial \psi}{\partial x}\,. \label{eq:defstreamslitTrans}
\end{eqnarray}
Eqs.~(\ref{eq:trans3a}) and (\ref{eq:trans3b}) turn to the inhomogeneous biharmonic equation:
\begin{eqnarray}
\left(\frac{\partial^2 }{\partial x^2}+\frac{\partial}{\partial y^2}\right)^2 \psi = \ra \frac{\partial \varphi_1}{\partial y}\,, \label{eq:FondaTrans}
\end{eqnarray}
with $\varphi_1$ given by Eq.~(\ref{eq:varphi12D}). 
The no-slip boundary conditions at the solid walls impose: 
\begin{eqnarray}
&&\frac{\partial \psi}{\partial y}(x,y=\pm\gamma/2)=\frac{\partial \psi}{\partial x}(x=0~\mathrm{and}~1,y)=\psi(x,y=\pm \gamma/2)=\psi(x=0~\mathrm{and}~1,y)=0\,.\label{eq:BCpsi2D}
\end{eqnarray}
It can be seen that the same equations  govern viscous flows induced by inhomogeneous temperature fields in a rectangular container, and we will use
the  method described in Refs.~\cite{Shankar1993,Shankar2002} to estimate the solution of Eq.~(\ref{eq:FondaTrans}).
In brief, the general solution of Eq.~(\ref{eq:FondaTrans}) is written as $\psi = \psi_{\mathrm{in}}+ \psi_{\mathrm{in}}$ with  $\psi_{\mathrm{in}}$ any solution of the inhomogeneous equation (i.e.\ with the right-hand term), and $\psi_{\mathrm{ho}}$ the solution of the homogenous biharmonic equation, so that $\psi$ fulfills the above boundary conditions. $\psi_{\mathrm{in}}$ can be found simply by a Fourier series expansion following $\varphi_1$, and
 $\psi_{\mathrm{ho}}$ can be found using a direct eigenfunction expansion, see below and Refs.~\cite{Shankar1993,Shankar2002} for details.  

For simplicity, we re-write $\varphi_1$ as:
\begin{eqnarray}
\varphi_1 = \ra \left(\frac{\partial \varphi_0}{\partial z}\right)^2 \sum_{n=1,\mathrm{odd}}^{\infty} \sum_{p=2,\mathrm{even}}^\infty h_{n,p}(y) \cos(n\pi x)\,, 
\end{eqnarray}
with $h_{n,p}(y)$ given in Eq.~(\ref{eq:varphi12D}). 
An inhomogeneous solution $\psi_{\mathrm{in}}$ of Eq.~(\ref{eq:FondaTrans}) can be easily found using the following Fourier series:
\begin{eqnarray}
\psi_{\mathrm{in}} = \left(\ra \frac{\partial \varphi_0}{\partial z}\right)^2 \sum_{n=1,\mathrm{odd}}^{\infty} \sum_{p=2,\mathrm{even}}^\infty a_{n,p}(y) \cos(n\pi x)\,,
\end{eqnarray}
with $a_{n,p}(y)$ solutions of the following ordinary differential equations:
\begin{eqnarray}
 a^{(4)}_{n,p}(y) - 2 \pi^2 n^2 a^{(2)}_{n,p}(y) + \pi^4 n^4  a_{n,p}(y) = h^{(1)}_{n,p}(y)\,.  
\end{eqnarray}
The solutions of these equations with the following boundary conditions:
\begin{eqnarray}
a^{(1)}_{n,p}(\pm \gamma/2) = a_{n,p}(\pm \gamma/2) = 0\,, 
\end{eqnarray}
lead to a Fourier series representing $\psi_{\mathrm{in}}$. For the sake of clarity, the functions $a_{n,p}(y)$ are not written here. 
This Fourier series fullfills all the boundary conditions given by Eq.~(\ref{eq:BCpsi2D}) except $\psi_{\mathrm{in}}(x=0,y) = -\psi_{\mathrm{in}}(x=1,y)  = \kappa(y) \neq  0$.

The next step consists therefore in finding the solution $\psi_{\mathrm{ho}}$ of the homogeneous biharmonic equation with the boundary conditions given by Eq.~(\ref{eq:BCpsi2D})  but with $\psi_{\mathrm{ho}}(x=0,y) = -\psi_{\mathrm{ho}}(x=1,y)  = -\kappa(y)$.
As the symmetry of the problem imposes that $\psi$ is an odd function of $y$, we will use the following odd eigunfunctions $\exp(\pm \lambda_n x) \phi_n(y)$ with:
\begin{eqnarray}
\phi_n(y) = y \cos(\lambda_n y) - \frac{\gamma}{2}\cot\left(\frac{\gamma \lambda_n}{2}\right)\sin(\lambda_n y)\,,
\end{eqnarray}
and $\lambda_n$  the complex roots of the transcendental equation:
\begin{eqnarray}
\sin(\lambda_n) = \lambda_n\,,
\end{eqnarray} 
which can be estimated by the Newton's method.
These eigenfunctions verify the  boundary conditions expected at $y = \pm \gamma/2$.
As $\psi$ is a real and odd function of $x$ with respect to $x=1/2$, an eigenfunction expansion for $\psi_{\mathrm{ho}}(x,y)$ is:
 \begin{eqnarray}
\psi_{\mathrm{ho}}(x,y) =\left(\ra \frac{\partial \varphi_0}{\partial z}\right)^2\sum_{n=1}^\infty \mathcal Re\left[ b_n \phi_n (y) \left(e^{-\lambda_n x} -e^{-\lambda_n (1-x)} \right)\right] \label{eq:sumpsishankar}
 \end{eqnarray}
where the complex numbers $b_n$ have to be determined from the boundary conditions at $x=0$ and $x=1$.
To get an approximate solution, we proceed as proposed by Shankar in Ref.~\cite{Shankar1993}.
First, the sum in Eq.~(\ref{eq:sumpsishankar}) is truncated to the first $N$ terms. 
Then, a least-squares procedure is used to find 
the coefficients $b_n$ which yield the best expected boundary conditions at $x=0$ for $m$ equidistant points over the interval $y=[0-\gamma/2]$. 
This procedure corresponding to the resolution of $2N$ linear algebraic equations is performed using Mathematica. 
The coefficients $b_n$ are rapidly converging and only a few eigenfunctions are needed to get an accurate estimate of $\psi_{\mathrm{ho}}(x,y)$. 

Figures~\ref{Fig:FigTransverse}(a) and  \ref{Fig:FigTransverse}(b) display the components $u_y$ and $u_x$ computed from the stream function $\psi = \psi_{\mathrm{in}} + \psi_{\mathrm{ho}}$ calculated using the above procedure for $\gamma = 1$. 
Although the velocity field ($u_x$,$u_y$) seem to suggest rotational symmetry for a square cross-section, this is not the case because of the gravity along $x$, and the maximum values of the components $u_x$ and $u_y$, although close, are not strictly equal.

%\bibliography{C:/Users/salmon/Documents/ARTICLE/BIB/onion} %You need to replace "rsc" on this line with the name of your .bib file


\end{document}